\documentclass[smallextended]{svjour3}       % onecolumn (second format)

\usepackage{siunitx}
\usepackage{graphicx}% Include figure files
\usepackage{dcolumn}% Align table columns on decimal point
\usepackage{bm}% bold math
\usepackage{hyperref}% add hypertext capabilities
\usepackage{subcaption}
\usepackage{float}
\usepackage{siunitx}%units
\usepackage{color}
\usepackage{amsmath}

\newcommand{\m}{\mathrm}
\newcommand{\RR}{\right}
\newcommand{\LL}{\left}

\newcommand{\eref}[1]{Eq.~(\ref{#1})}
\newcommand{\fref}[1]{Fig.~\ref{#1}}

\journalname{Journal of Low Temperature Physics}

\begin{document}

%\preprint{APS/123-QED}

\title{High precision displacement sensing of monolithic piezoelectric disk resonators using a single-electron transistor}

\author{J. Li         \and
        J. T. Santos \and
        M. A. Sillanp\"a\"a
}

 %\author{Jian Li}
 %\affiliation{Interdisciplinary Center of Quantum Information, College of Science, National University of Defense Technology, Changsha 410073, China}

 %\author{Jorge T. Santos}
 %\affiliation{Department of Applied Physics, Aalto University, P.O. Box 15100, FI-00076 AALTO, Finland}%

 %\author{Mika A. Sillanp\"a\"a}
 %\affiliation{Department of Applied Physics, Aalto University, P.O. Box 15100, FI-00076 AALTO, Finland}%

\institute{J. Li \at
              Interdisciplinary Center of Quantum Information, National University of Defense Technology, Changsha 410073, China\\
              \email{lijian16@nudt.edu.cn}           %  \\
           \and
           J. T. Santos \at
              Department of Applied Physics, Aalto University, P.O. Box 15100, FI-00076 AALTO, Finland
              \and
           M. A. Sillanp\"a\"a \at
              Department of Applied Physics, Aalto University, P.O. Box 15100, FI-00076 AALTO, Finland
}

\date{Received: date / Accepted: date}

\maketitle

\begin{abstract}
\noindent A single electron transistor (SET) can be used as an extremely sensitive charge detector. Mechanical displacements can be converted into charge, and hence SETs can become sensitive detectors of mechanical oscillations. For studying small-energy oscillations, an important approach to realize the mechanical resonators is to use piezoelectric materials. Besides coupling to traditional electric circuitry, the strain-generated piezoelectric charge allows for measuring ultrasmall oscillations via SET detection. Here we explore the usage of SETs to detect the shear mode oscillations of a \SI{6}{\milli\meter} diameter quartz disk resonator with a resonance frequency around \SI{9}{\mega\hertz}. We measure the mechanical oscillations using either a conventional dc SET, or use  the SET as a homodyne or heterodyne mixer, or finally, as a radio-frequency single-electron transistor (RF-SET). The RF-SET readout is shown to be the most sensitive method, allowing us to measure mechanical displacement amplitudes below $10^{-13}$ meters. We conclude that a detection based on a SET offers a potential to reach the sensitivity at the quantum limit of the mechanical vibrations.
\keywords{Single Electron Transistor \and Piezoelectric Resonators \and Displacement Detectors}
\end{abstract}

\section{\label{sec:level1}Introduction}

The measurement of atto-Newton forces, or sub-\r{A}ngstrom displacements, has a wide range of existing or potential applications, from magnetic resonance force microscopy \cite{slides} to the study of frontiers of physics such as gravitational waves or quantum effects in the motion of mechanical resonators \cite{Palomaki710,Wollman952,JuhaSqueeze,Lecocq2015}. Magnetomotive \cite{Cleland1996,Cleland1999256,Mohanty2000,Pashkin08}, optical interferometric \cite{Carr1999,Hakseong2017} as well as mixing \cite{vanderZant2009,Bachtold2009} detection techniques have been widely utilized in fundamental research. Moreover, we mention approaches that base on the concept of cavity optomechanics, where the mechanical oscillations affect the electromagnetic fields inside either an optical \cite{Cohadon1999,Aspelmeyer2006cool,Heidmann2006} or an electrical \cite{Lehnert2008Nph} resonator.

The original detection approaches used in fundamental research included, for instance,  the magnetomotive technique that however, suffers from the need of high fields and bulky equipment. Similarly, optical interferometry also endures the latter issue, and the laser spot size limits the size of the mechanical resonator that can be studied. While impressive quantum experiments have been performed using cavity optomechanics either in optical or in microwave frequency range, the laser or microwave powers that are needed tend to be high and heat up the system. Therefore, one should look for even less intrusive detection techniques that would work at extremely small power. To this end, several experiments have taken advantage of a single electron transistor (SET) \cite{Flees1997,Joyez1994}, which is an ultimate charge sensitive device, and can be capacitively coupled to mechanical devices \cite{Knobel2003,LaHaye74,Nakamura2010nems}. In such scheme, a dc voltage bias of the order of several volts is applied to a conductive mechanical resonator, and the mechanical vibrations modulate the amount of charge coupled to the SET. In the most delicate measurements, the SET has been replaced by superconducting qubits that bring the system in the quantum regime \cite{LaHaye2009,ClelandMartinis,transmonnems,LSETNEMSexp,Delsing2014,LaHaye2016}.

%The increase of the mechanical element biasing voltage increases  the amount of charge coupled to the SET island, improving the sensitivity of the displacement detection. However, higher biasing voltages also enlarge the back-action on the mechanical resonator by the SET, thus limiting the displacement sensitivity of this scheme to \SI{4e-16}{\meter\per\sqrt\hertz} for a model cantilever \cite{Armour2002}.

In our recent work \cite{Santos2017}, we show how a massive  piezoelectric resonator coupled to superconducting microcircuits \cite{woolley2016} can be analyzed as a cavity optomechanical device, with implications in studies of quantum-mechanical phenomena. 
%Interestingly, piezoelectricity is naturally compatible with SET detection, since if the mechanical resonator is made of a piezoelectric material, the SET can sensitively observe the strain-induced piezoelectric charge  \cite{Knobel2003,Knobel2002,Sidles91}. 
The sensitivity advantage offered by piezoelectric transduction \cite{Knobel2003,Knobel2002,Sidles91} is highly beneficial for studying and further using the quantum properties of truly macroscopic resonators that can exhibit long energy lifetimes but nearly vanishing amplitudes of their  zero-point motion.

In practical usage, a SET has a very limited bandwidth. The typical large tunnel junction resistance needed for the Coulomb blockade, together with the ever present stray capacitance from the cables limits the output bandwidth to about \SI{e4}{\hertz}. A first stage amplifier in close proximity to the SET \cite{Pettersson96,Visscher96} can extend the bandwidth to $10^6 ... 10^7$ Hz at best. There are two standard options to circumvent this limitation and broaden the spectral range of the SET. One is to use it as a radio-frequency (RF) mixer, either in homodyne, or in heterodyne measurement scheme \cite{knobel81,Abuelmaatti2009}, taking advantage of the nonlinear dependence of the SET current response to the gate charge. This approach does not increase the instantaneous bandwidth of the SET, which is still limited by the RC charging time, but allows to tune the measurement center frequency up to \SI{10}{\giga\hertz} \cite{knobel81}. The second approach is to use an RF-SET \cite{Schoelkopf1238}, where a series inductance resonates with the cable stray capacitance, creating an LC tank resonator that impedance matches the SET to the low impedance cables, unlocking bandwidth up to \SI{100}{\mega\hertz} \cite{Devoret2000}.

\begin{figure*} [t]
\centering
    \includegraphics[width=0.8\textwidth]{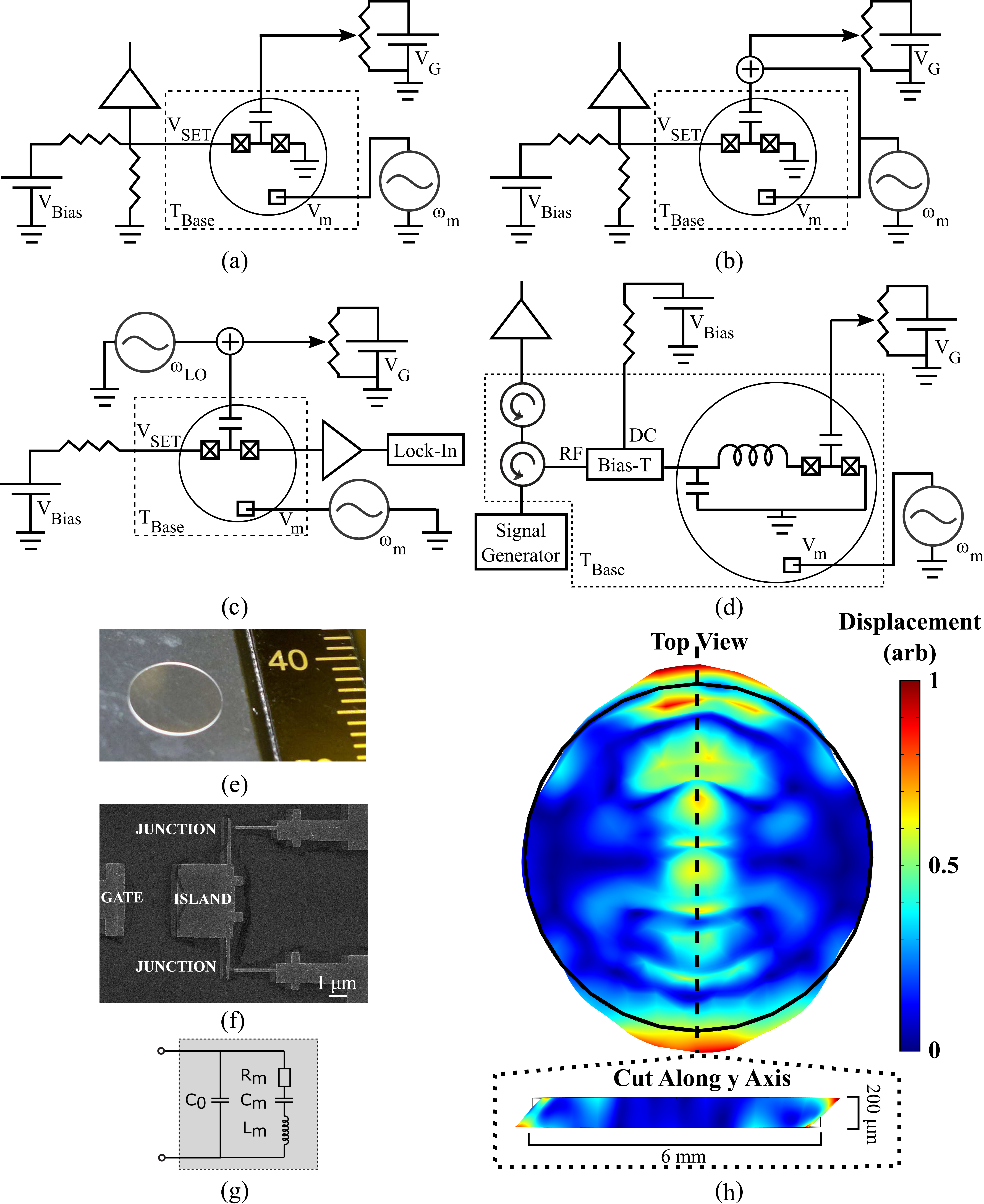}
  \caption{Measurement scheme for detecting vibrations of the massive quartz disk via SET, using (a) DC rectification; (b) homodyne and (c) heterodyne detection using the SET as a mixer; and (d) RF-SET. The  disk represent the quartz resonator/substrate. Drawings not to scale. (e) Photograph of the \SI{6}{\milli\meter} quartz disks used both as the mechanical element and as the substrate where the SET circuit is fabricated. (f) Scanning Electron Microscope (SEM) image of the SET device with a \SI{3}{\micro\meter} x \SI{4}{\micro\meter} island. (g)  Electrical resonant circuit equivalent to a piezoelectric mechanical resonator. (h) Simulated shear mode shape. The black circle represents the quartz disk edges, and the color codes indicate the deformation. The displacement values are normalized to the maximum displacement. The side view dimensions are not in scale.}
  \label{fig:systems}
\end{figure*}

In this article we investigate a mechanical system that is a 6 millimetre diameter and 200 microns thick quartz disk resonator (Fig.~\ref{fig:systems}e) that has the first shear mode resonance frequency at $\omega_0/2\pi \sim $ \SI{9}{\mega\hertz}. A major motivation for studying such monolithic resonators is that they can have exceedingly high mechanical quality factors, paving the way towards macroscopic quantum mechanics \cite{goryachev2012}. These systems are truly macroscopic with the mode effective mass around 20 mg. We study the prospect of measuring the disk vibrations using four different SET detection setups: DC (Fig.~\ref{fig:systems}a), homodyne mixing (Fig.~\ref{fig:systems}b), heterodyne  mixing (Fig.~\ref{fig:systems}c), and RF-SET (Fig.~\ref{fig:systems}d). The SETs are fabricated directly on top of the quartz disk (Fig.~\ref{fig:systems}f), that acts  as both the mechanical element and the substrate for the fabrication of the electrical circuit. The basic idea  is independent of the measurement scheme chosen: the deformation of the quartz disk generates piezoelectric charge on the chip surface; the charge created in the vicinity of the SET island couples to it, thereby modulating the tunnelling current or the SET resistance.

\section{\label{sec:level2}Piezoelectric Resonator and SET}

%A piezoelectric resonator can be represented by an equivalent series RLC circuit like the one in Fig.~\ref{fig:systems}g. Let us consider a circular quartz disk with the parameters shown on Table \ref{table} and their corresponding  values. The geometric capacitance in the plate-capacitor approximation is $C_0=\epsilon_0 \epsilon_r A / z$ and the other equivalent circuit parameters can be calculated as $C_m=K_0^2 C_0$, $L_m=[(2 \pi \omega_m)^2 C_m]^{-1}$ and $R_m=(C_m \omega_m Q)^{-1}$, where $Q$ is the mechanical quality factor. 

A piezoelectric resonator can be represented by an equivalent series RLC  circuit like the one in Fig.~\ref{fig:systems}g. Let us consider a circular quartz disk with the surface area $A$, thickness $z$, shear mode stress coefficient $e_{s}$, shear modulus $Y_s$, relative permittivity $\epsilon_r$, and piezoelectric coupling coefficient $K_0^2=e_{s}^2/{(\epsilon_0 \epsilon_r Y_s)}$. The geometric capacitance in the plate-capacitor approximation is $C_0=\epsilon_0 \epsilon_r A / z$ and the other equivalent circuit parameters can be calculated as $C_m=K_0^2 C_0$, $L_m=[(2 \pi \omega_0)^2 C_m]^{-1}$ and $R_m=(C_m \omega_0 Q)^{-1}$, where $Q$ is the mechanical quality factor. In Table \ref{table}, we summarize the quantities mentioned above.

\begin{table}[]
\caption{Geometric and material parameters of the quartz disk resonator and an approximation of the correspondent equivalent electrical quantities when modeling the mechanical resonator  as the RLC circuit of \fref{fig:systems}g. }
%\centering
\begin{tabular}{llll|}
\hline Parameter                       &Symbol       &    Value        &   Unit      \\
\hline
Disk Surface Area                           &$A$             & \num{2.8e-5}     &\si{\square\meter}               \\
Disk Thickness                                    &$z$          & \num{2e-4}         &\si{\meter}            \\
Quartz Shear Modulus                     &$Y_s$           & \num{3e10}        &\si{\pascal}       \\
Quartz Shear Stress Coefficient           &$e_s$        & \num{1e-1}         &\si{\coulomb\per\square\meter}      \\
Quartz Relative Permittivity                 &$\epsilon_r$     & \num{4.5}       &             \\
Disk Mechanical Frequency                &$\omega_0$     & \num{2\pi x 9}                       &\si{\mega\hertz}                      \\
Coupling Coefficient & $K_0^2=e_{s}^2/{(\epsilon_0 \epsilon_r Y_s)}$         & \num{8e-3}     & \si{\square\meter\per\coulomb}  \\
\hline
Geometric capacitance &$C_0=\epsilon_0 \epsilon_r A / z$      &\num{5e-12}  & \si{\farad}        \\
Mechanical capacitance & $C_m=K_0^2 C_0$      &\num{4e-14}  & \si{\farad}       \\
Mechanical inductance &$L_m=[\omega_m^2 C_m]^{-1}$        &\num{8e-3}   & \si{\henry}       \\
Mechanical resistance & $R_m=(C_m \omega_m Q)^{-1}$                     &\num{30}  & \si{\ohm} \\
Mechanical quality factor & $Q$       &\num{15e3}  & \\
\hline
\end{tabular}
\label{table}
\end{table}

A static shear deformation $\Delta x_0$ corresponding to an applied potential difference $V$ between the two faces of the quartz disk is given as
\begin{equation} \label{eq:displacementStatic}
\Delta x_0= \frac{\epsilon_r \epsilon_0}{e_s} V \,,
%\Delta x=d_{14} V \approx \num{0.7e-12} V (\si{\meter})
\end{equation}
that amounts to less than a nanometer displacement per volt. In a dynamical situation, close to the mechanical resonance, the resonator amplifies the applied voltage $V_m$ with the frequency $\omega_m$, cf.~the equivalent electrical circuit in  Fig.~\ref{fig:systems}g, and one obtains the dynamical vibration amplitude
\begin{equation} 
\label{eq:displacement}
\Delta x= \frac{e_s V_m}{\gamma Y_s z} \frac{\sinh (\gamma z /2) }{\cosh (\gamma z /2)}  \,,
\end{equation}
where
\begin{equation} 
%\label{eq:displacement}
\gamma^2 = -\frac{\omega_m^2/\nu_0^2}{1+i /Q}  \,,
\end{equation}
and $\nu_0 \approx 3540$ ms$^{-1}$ is the sound speed of shear waves in quartz.

In the experiment, we actuate the vibrations through a separate excitation electrode shown in  \fref{fig:systems}a-d. Equation (\ref{eq:displacement}) holds only for the part of the chip covered by the excitation electrode, while we are interested in the amplitude under the  SET island somewhat distant from the electrode. In order to enable experimental comparison, we need to know how much the amplitude is reduced from that in \eref{eq:displacement} due to the small overlap. To this end we run a Comsol simulation and compare a full electrode coverage versus the actual chip layout, obtaining a reduction factor $\beta \simeq 0.11$ that should multiply the right-hand-side of \eref{eq:displacement}.

Assuming a uniform piezoelectric charge distribution across the disk surface, a shear deformation $\Delta x$ corresponds to a shear strain $\lambda_s = \Delta x/z$ and generates a piezoelectric surface charge density $\sigma_q = \lambda_s e_{s}$. Then, the number of electron charges coupled to a SET island  is
\begin{equation} \label{eq:charge}
n_{\m{isl}}\approx \sigma_q A_{\m{isl}}=\frac{\Delta x e_{s} A_{\m{isl}}}{z e} \,.
\end{equation}
%
%The potential difference between the disk surfaces generated due to the applied strain can be calculated from $V_s\approx \sigma_q A / C_0$.
Here,  $A_{\m{isl}}$ is the effective  quartz surface area over which the SET sees the piezo charge. It is of the order the  SET island area, but larger by a small numerical factor.

 We model the SET response to charge as follows. If there is a total charge $n_g$ coupled to the SET island, we suppose the periodic gate charge response of the SET can be approximated as sinusoidal:
\begin{equation} \label{eq:setsine}
%I_{\m{SET}} = \frac{I_0}{2} \LL[ 1-\cos \LL(\frac{\pi n_g}{2} \RR)\RR] + i_0\,,
I_{\m{SET}} =  i_0 - \frac{I_0}{2} \cos \LL(\pi n_g\RR) \,,
\end{equation}
which represents the $2e$ periodicity of a superconducting SET. Here, $I_0$ is the peak-to-peak amplitude of gate modulation, and $i_0$ the average current, both of which  depend on the SET bias. The nonlinearity of the SET is the basis for the DC and mixing detection methods. The most general case of gate charge setting discussed below consists of three terms. There is usually a non-zero constant offset $n_{g0}$, and around the offset the gate charge is supposed to oscillate sinusoidally in time due to the mechanical oscillations driven at the frequency $\omega_m$. Moreover, in the mixing schemes, there is a strong local oscillator at the frequency $\omega_{\m{LO}}$ and with the amplitude $n_{\m{LO}}$:
\begin{equation}
 \label{eq:sinetime}
n_g(t) = n_{g0} + n_{\m{isl}} \cos \LL(\omega_m t\RR) + n_{\m{LO}} \cos \LL(\omega_{\m{LO}} t\RR) \,.
\end{equation}
The nested sinusoidal behaviour of Eqs.~(\ref{eq:setsine},\ref{eq:sinetime}) will be seen in the experimental data below.

\section{\label{sec:results}Results}

%Issues:
%{\color{red} NO INFO ON MEAS BANDWIDTHS, ETC, HOW TO CALCULATE SENSITIVITIES??????}
%{\color{red} transconductance gm calculated from the Iset vs Vmech  for:  dQ/sqrt(df) = sqrt(S(0)/gm = sqrt(2eI) sqrt(SI(0)/gm}
%{\color{red} $http://folk.uio.no/yurig/Publications/Before_96/14a.pdf$}

In the measurements, we used two individual samples, labeled A and B. Sample A was used in all the other measurements except in the heterodyne scheme (\fref{fig:systems}c). The samples were essentially similar, but sample B had a smaller junction resistance, although this plays only a minor quantitative role in the results.

We begin with discussing the current-voltage (IV) properties of the SET in sample A. It comprises a \SI{3}{\micro\meter} $\times$ \SI{4}{\micro\meter} island delimited by two Al (\SI{30}{\nano\meter}) / AlOx / Al (\SI{60}{\nano\meter}) tunnel junctions made  by shadow evaporation. For the measurements, the edges of the chip with the SET are firmly glued to a sample stage, but the center of the chip is free to vibrate. All the measurements discussed in this paper were carried out in a dilution refrigerator at a temperature of $\sim 20$ mK. The sum resistance of the two junctions of our SET in sample A is approximately $\SI{169}{\kilo\ohm}$, and the charging energy is estimated as 0.3 K. 
%$R_{J1}+R_{J2} \approx \SI{169}{\kilo\ohm}$. 

%The current supplied by the bias voltage source for a voltage $V_{SET Bias}$ is approximately $V_{SET Bias}/\SI{8.8}{\mega\ohm}$, because the \SI{8.8}{\mega\ohm} resistor is much larger than the parallel resistance of the SET with the \SI{6.8}{\kilo\ohm} resistor. Then, the source-drain current can be calculated from
%\begin{equation} \label{eq:ISET}
%I_{SET}=\frac{V_{SET Bias}}{\SI{8.8}{\mega\ohm}}-\frac{V_{SET}}{\SI{6.8}{\kilo\ohm}}
%\end{equation}
%where $V_{SET}$ is the source-drain voltage on the SET, measured by the digital voltimeter.

%
\begin{figure*}[t]
        \begin{subfigure}[2]{0.333\textwidth}
                \includegraphics[width=\linewidth]{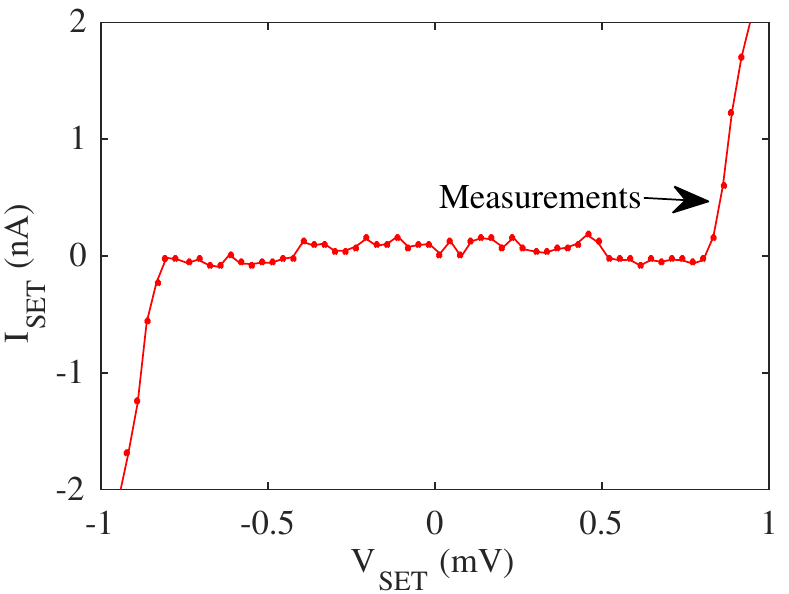}
                \caption{}
                \label{fig:2}
        \end{subfigure}%
        \begin{subfigure}[3]{0.333\textwidth}
                \includegraphics[width=\linewidth]{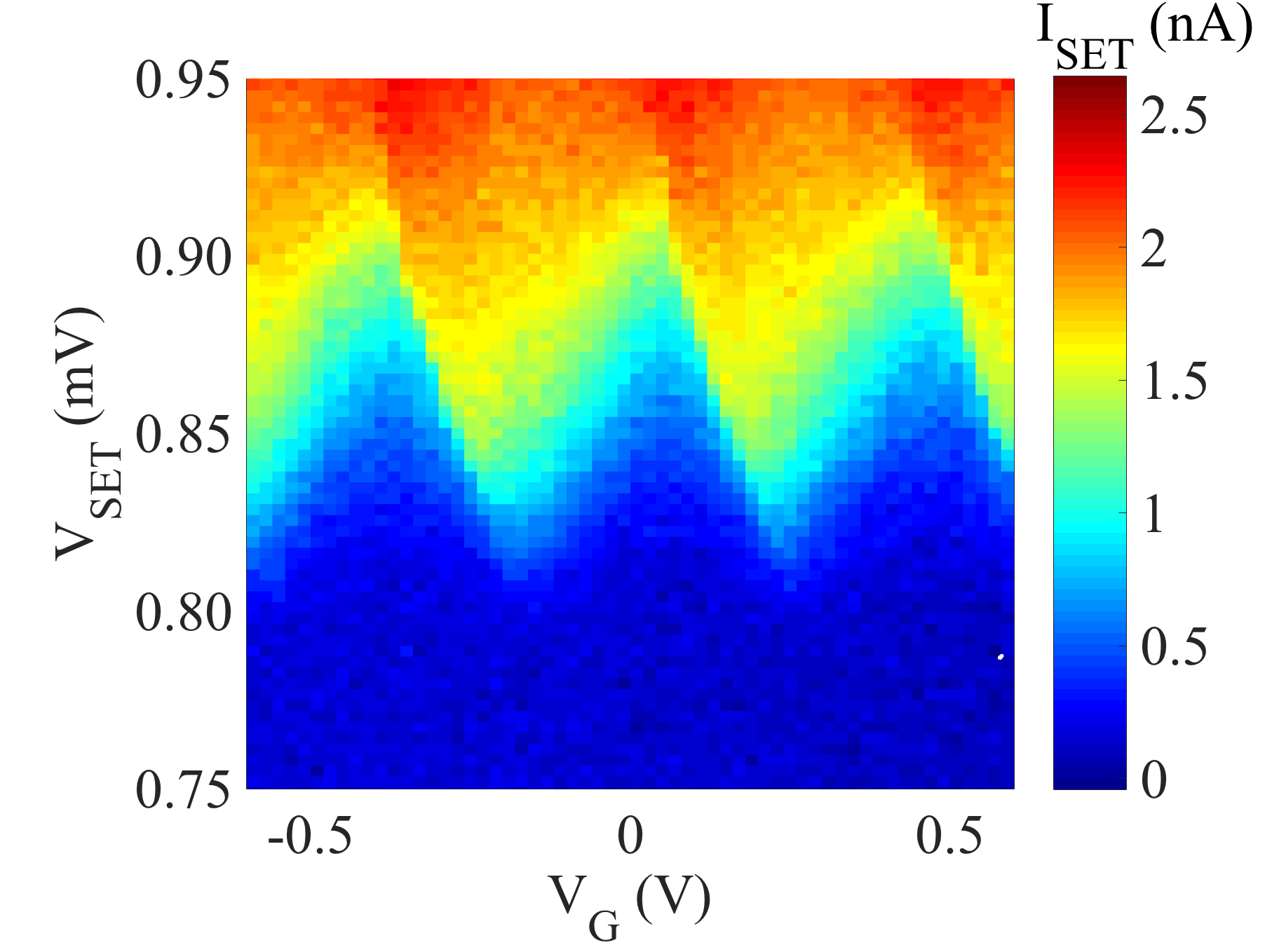}
                \caption{}
                \label{fig:3}
        \end{subfigure}%
        \begin{subfigure}[4]{0.333\textwidth}
                \includegraphics[width=\linewidth]{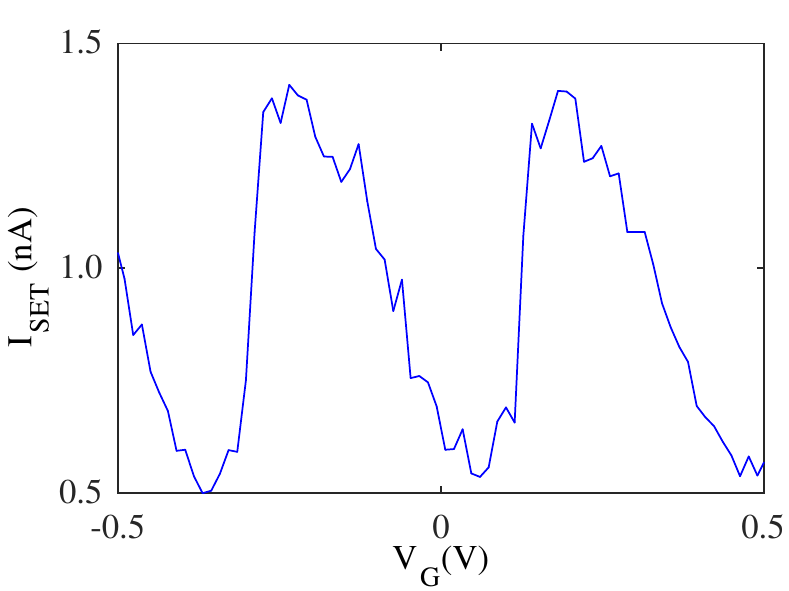}
                \caption{}
                \label{fig:4}
        \end{subfigure}%
        \caption{(\subref{fig:2}) Sample A: DC IV curve for the SET measured with the setup of Fig.~\ref{fig:systems}a when the gate voltage $V_G=0$. (\subref{fig:3}) Dependence of the current through the SET  on $V_G$ for the bias values around the steepest part of the IV curve presented in Fig. \ref{fig:2}. (\subref{fig:4}) Dependence of the SET current on gate voltage for an optimal $V_{\m{SET}}$ biasing of \SI{0.86}{\milli\volt}.}\label{fig:IV}
\end{figure*}

The standard DC IV measurement involves the biasing scheme shown to the left in Figs.~\ref{fig:systems}a-c. We apply current bias through a \SI{8.8}{\mega\ohm} resistor for the schemes in Figs.~\ref{fig:systems}a-b and through \SI{100}{\kilo\ohm} in \fref{fig:systems}c. Figure \ref{fig:2} shows the DC IV properties for our SET device. The width of the $I_{\m{SET}}$ current plateau is mostly set by the superconductor gap in our superconducting device. The Coulomb blockade modulation as a function of the gate voltage $V_G$ is visible at edges of the plateau.

In the quartz vibration measurements with sample A, we use the optimal $V_{\m{SET}} \approx \SI{0.86}{\milli\volt}$ biasing point around the gap edge, which maximizes the modulation of $I_{\m{SET}}$ by $V_G$, marked by an arrow in \fref{fig:2}. Figure \ref{fig:4} displays the gate dependence of the current around such optimal biasing, showing approximately sinusoidal dependence with a peak-to-peak amplitude of $I_0 \approx $ \SI{0.9}{\nano\ampere}. Here and throughout the paper we refer to the gate voltage as the value at the room temperature generator, preceding the cryogenic voltage division. 

%The gate period is around \SI{0.4}{\volt} We choose $V_{SET} \approx \SI{-0.23}{\milli\volt}$ to bias the device because at such point the $I_{\m{SET}}$ dependence on $V_G$ is approximately sinusoidal with maximal amplitude. Such dependence is plotted in Fig. \ref{fig:4}, where $I_{\m{SET}}$ has an average value around \SI{159}{\nano\ampere} and a peak to peak amplitude of \SI{0.9}{\nano\ampere}.

%In a superconductive SET the gate period corresponds to the charge of two electrons, so the gate capacitance can be calculated from $Q_e = C_G V_{per}$, where $Q_e$ is the elementary charge, $C_G$ the gate capacitance and $V_{per}=\SI{0.54}{\milli\volt}$ the gate period observed in Fig.\ref{fig:4} after applying the attenuation along the feed line. So, the gate capacitance of our SET is $C_G=\SI{0.6}{\femto\farad}$.

\subsection{\label{sec:dc}DC readout with the SET}   

%multimeter averages, DC offset?, since piezo only creates + or - charge for same signal V, the offset change even for signal out of band???
We begin with a simple  rectification approach where $n_{\m{LO}} = 0$ in \eref{eq:sinetime}, allowing to measure large-amplitude vibrations. Equation (\ref{eq:setsine}) becomes 
%
%\begin{equation} 
%\label{eq:setdcbessel}
%\begin{split}
%I &=  \frac{I_0}{2} - \frac{I_0}{2}  \cos \LL(\frac{\pi}{2} (n_{g0} + n_{\m{isl}} \cos \LL(\omega t\RR) ) \RR)  \\
%& =   \frac{I_0}{2}  -\frac{I_0}{2}  \cos \LL(\frac{\pi}{2} n_{g0} \RR) \cos \LL(\frac{\pi}{2} n_{\m{isl}} \cos \LL(\omega t\RR)  \RR) +\\
%& + \frac{I_0}{2}  \sin \LL(\frac{\pi}{2} n_{g0} \RR) \sin \LL(\frac{\pi}{2} n_{\m{isl}} \cos \LL(\omega t\RR)  \RR) = \\
%I_{\m{SET}} &= \frac{I_0}{2}\LL[1  - \cos \LL(\frac{\pi}{2} n_{g0} \RR) J_0 \LL(\frac{\pi}{2} n_{\m{isl}}   \RR) \RR] +i_0 \,.
%\end{split}
%\end{equation}
%
%
\begin{equation} 
\label{eq:setdcbessel}
\begin{split}
%I &= i_0 - \frac{I_0}{2}  \cos \LL(\frac{\pi}{2} (n_{g0} + n_{\m{isl}} \cos \LL(\omega t\RR) ) \RR)  \\
%& =   i_0  -\frac{I_0}{2}  \cos \LL(\frac{\pi}{2} n_{g0} \RR) \cos \LL(\frac{\pi}{2} n_{\m{isl}} \cos \LL(\omega t\RR)  \RR) +\\
%& + \frac{I_0}{2}  \sin \LL(\frac{\pi}{2} n_{g0} \RR) \sin \LL(\frac{\pi}{2} n_{\m{isl}} \cos \LL(\omega t\RR)  \RR) = \\
%I_{\m{SET}} &= \frac{I_0}{2}\LL[1  - \cos \LL(\frac{\pi}{2} n_{g0} \RR) J_0 \LL(\frac{\pi}{2} n_{\m{isl}}   \RR) \RR] +i_0 \,.
I_{\m{SET}} &=  i_0 - \frac{I_0}{2} \cos \LL(\pi n_{g0} \RR) J_0 \LL(\pi n_{\m{isl}}   \RR)  \,,
\end{split}
\end{equation}
where $J_0$ is the zeroth order Bessel function of the first kind. 

\begin{figure*}
        \begin{subfigure}[6]{0.333\textwidth}
                \includegraphics[width=\linewidth]{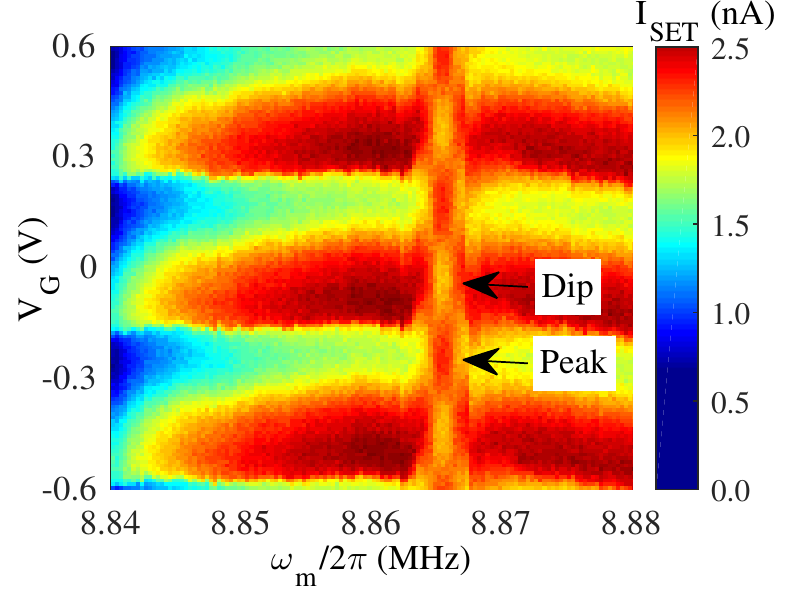}
                \caption{}
                \label{fig:6}
        \end{subfigure}%
        \begin{subfigure}[5]{0.333\textwidth}
                \includegraphics[width=\linewidth]{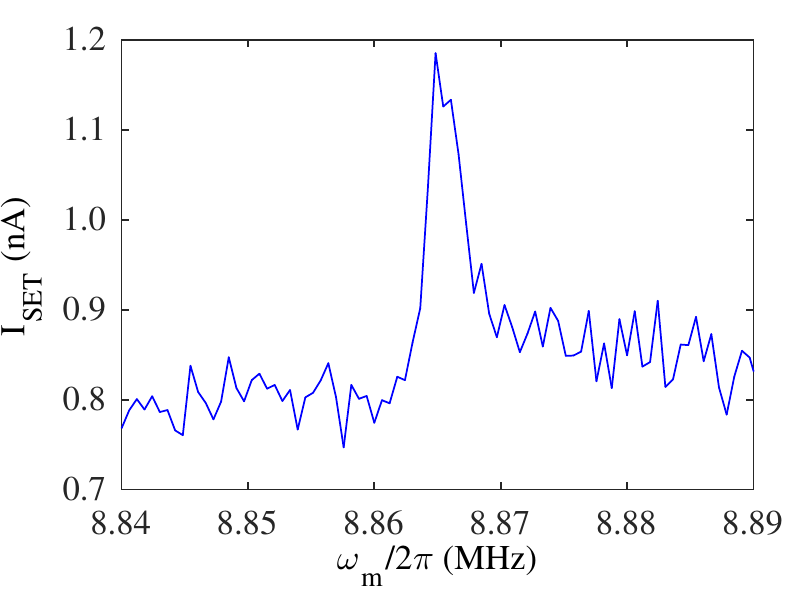}
                \caption{}
                \label{fig:5}
        \end{subfigure}%
        \begin{subfigure}[8]{0.333\textwidth}
                \includegraphics[width=\linewidth]{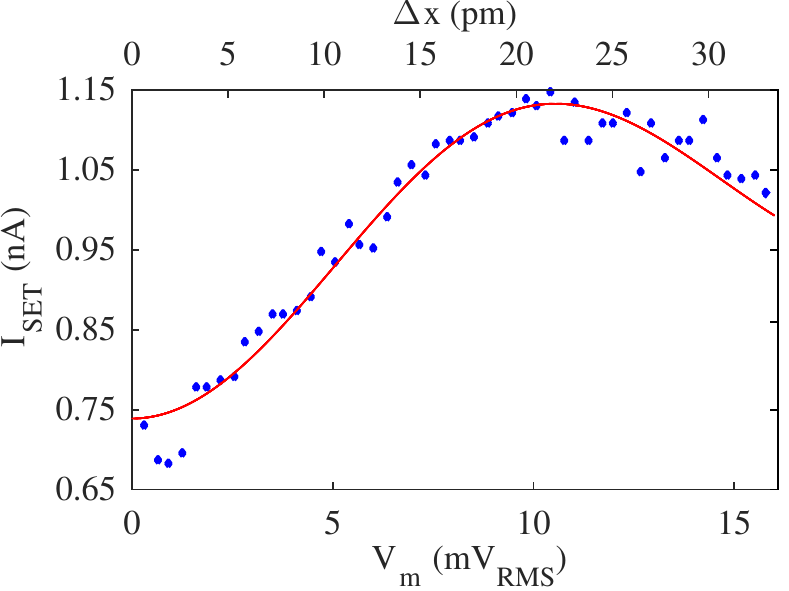}
                \caption{}
                \label{fig:8}
        \end{subfigure}%
        \caption{Response of the DC measurement SET setup of  \fref{fig:systems}a (sample A), biased with $V_{SET}=\SI{0.86}{\milli\volt}$,  to the driven motion in the piezoelectric quartz disk resonator: (\subref{fig:6}) Gate voltage modulation of the SET current $I_{\m{SET}}$ around the mechanical mode frequency at $\approx \SI{8.865}{\mega\hertz}$ when $V_{m} \approx 11$ mV$_{\m{RMS}}$; (\subref{fig:5}) The mechanical resonance peak for an actuation voltage $V_{m} \approx 13$ mV$_{\m{RMS}}$, when $V_G \simeq 0.12$ V; (\subref{fig:8}) The SET current when the mechanical resonator is excited on-resonance at increasing excitation voltages. The solid line is a theoretical curve based on Eqs.~(\ref{eq:displacement},\ref{eq:charge},\ref{eq:setdcbessel}) with $n_{g0}= 0.31$ and an island interaction area $A_{\m{isl}}\approx 18 \, (\SI{}{\micro\meter})^2$.}
        \label{fig:rect}
\end{figure*}

Now we excite the mechanical vibrations with voltage, thereby creating non-zero $n_{\m{isl}} $. Physically, the piezoelectric charge modulates the SET current by changing the potential of the island, in the same way as the gate voltage. In \fref{fig:6} we display  such a measurement for different values of  $V_G$. The mechanical resonance can be observed at $f_{0}=\SI{8.865}{\mega\hertz}$, as a sharp feature that depends on the gate bias.  Looking at \eref{eq:setdcbessel}, the data qualitatively matches the prediction. At the current minima or maxima, where $\cos \LL(\pi n_{g0} \RR) = 1$, the peak is roughly the maximum in amplitude, but disappears at the intermediate values when $\cos \LL(\pi n_{g0}  \RR) \approx 0$. Also, the sign of the peak changes according to the prediction.

In \fref{fig:5} we display an example of a resonance peak measured at a rather large mechanical excitation voltage of $\SI{13}{\milli\volt}_{\m{RMS}}$, corresponding to a displacement around \SI{27e-12}{\meter} calculated from  \eref{eq:displacement}. At higher mechanical excitation amplitudes the SET current, however, starts to decrease as seen in \fref{fig:8} at $V_{m}\approx \SI{10}{\milli\volt}_{\m{RMS}}$ where the driven charge covers more than one gate period. We can make an independent estimate of the cross-over voltage by the fact that it corresponds to the first minimum of $J_0 \LL(\pi n_{\m{isl}}\RR)$, at $n_{\m{isl}}\approx \SI{1.2}{\elementarycharge}$, and using Eqs.~(\ref{eq:displacement},\ref{eq:charge}) yielding $V_{m}\approx \SI{11}{\milli\volt}$, in a good agreement with the measurement.

\begin{figure*}
        \begin{subfigure}[6]{0.45\textwidth}
                \includegraphics[width=\linewidth]{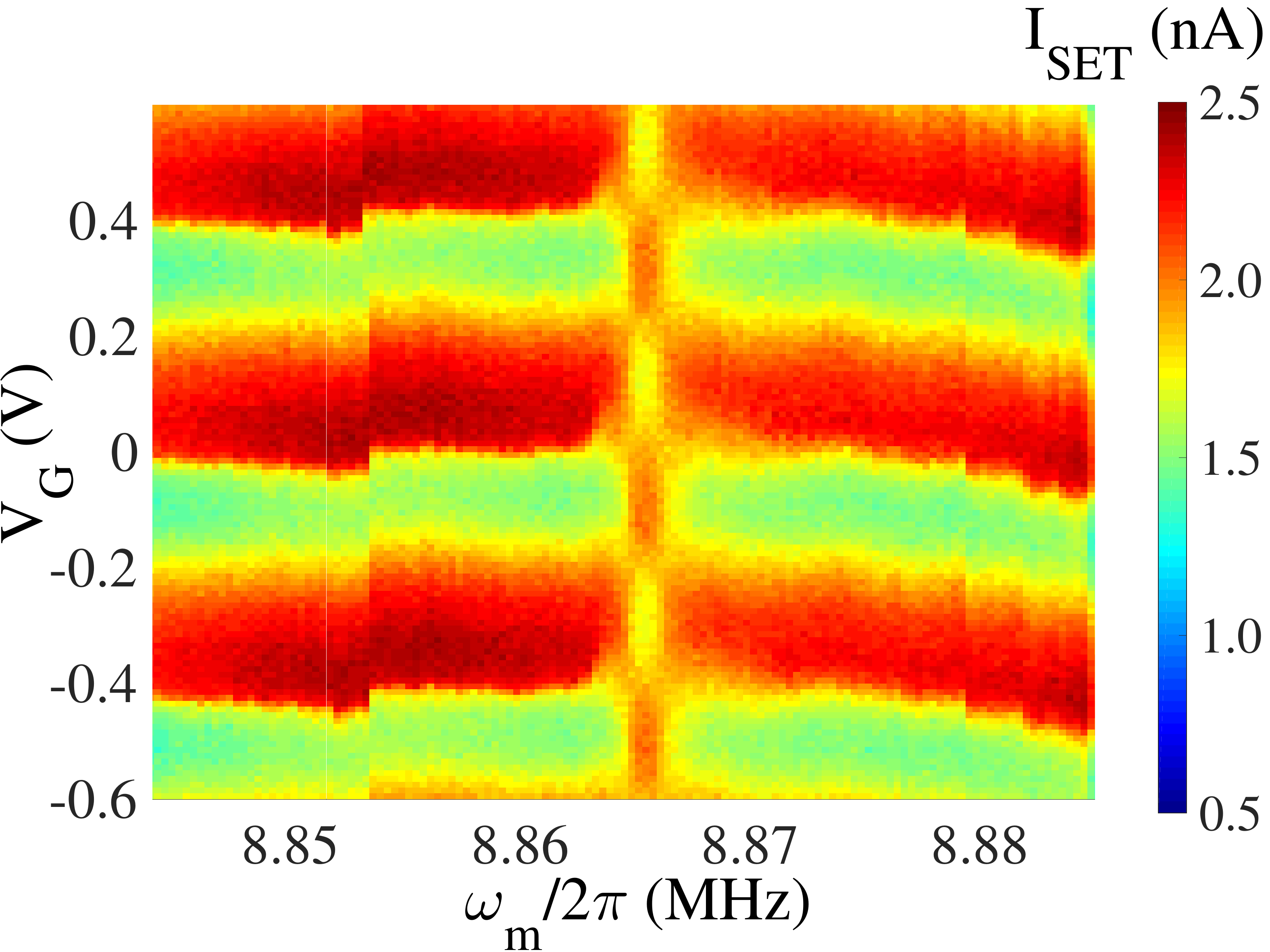}
                \caption{}
                \label{fig:9}
        \end{subfigure}%
        \begin{subfigure}[7]{0.45\textwidth}
                \includegraphics[width=\linewidth]{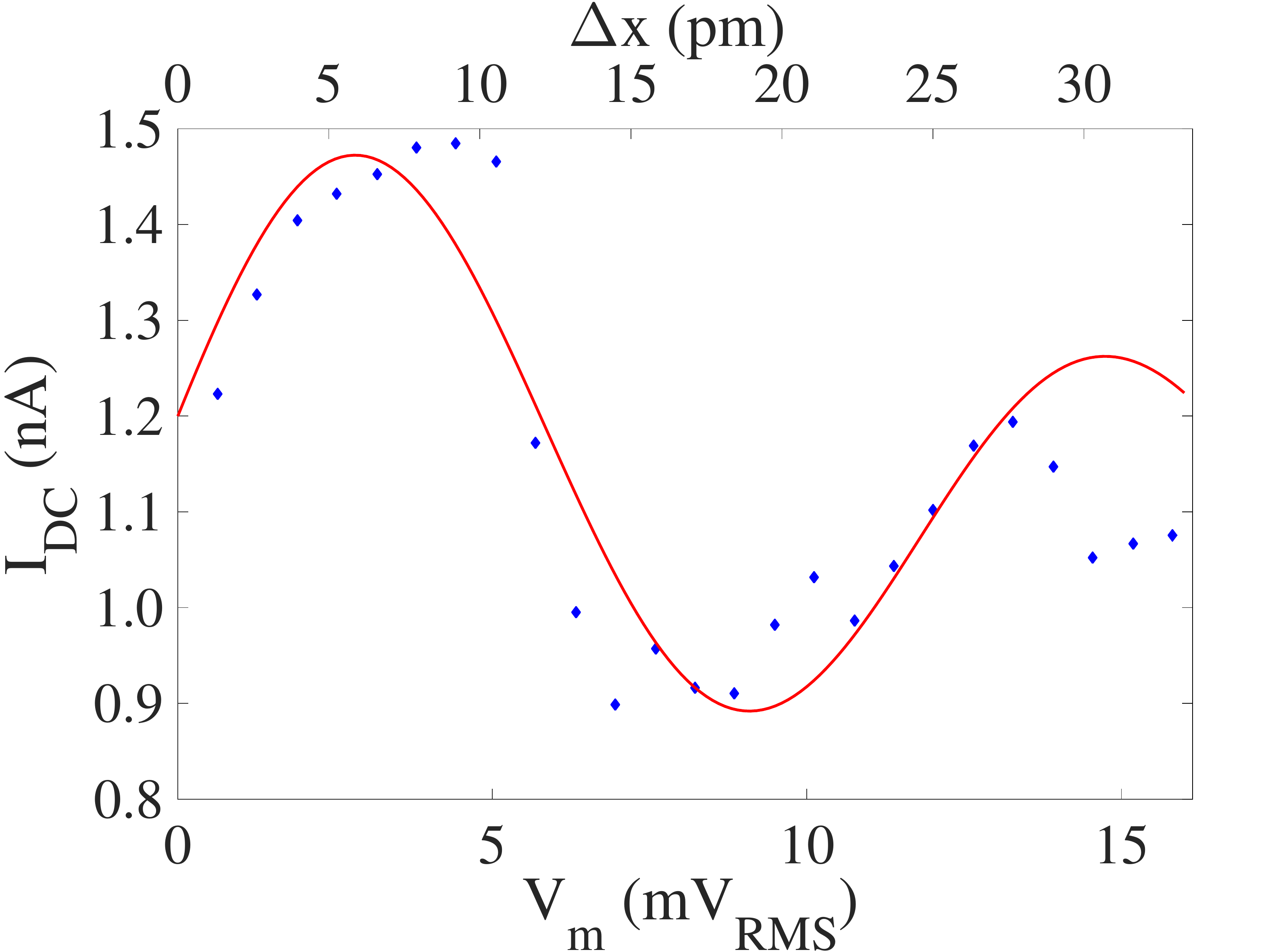}
                \caption{}
                \label{fig:11}
        \end{subfigure}%
        \caption{Quartz disk vibrations detected using the homodyne detection scheme in \fref{fig:systems}b (sample A, with the SET biased as shown in \fref{fig:2}):  (\subref{fig:9}) The SET current as a function of gate voltage and actuation frequency when $V_{m}=\SI{2.6}{\milli\volt}$; (\subref{fig:11}) SET drain to source current when the mechanical resonator is excited at its resonance frequency $f_{0}=\SI{8.865}{\mega\hertz}$ with different excitation amplitudes. The device is biased with $V_G=\SI{0.3}{\volt}$. The solid line is a theoretical curve based on Eqs. (\ref{eq:displacement},\ref{eq:charge},\ref{eq:sethomobessel}) with the same parameters used in \fref{fig:8} except  $n_{g0}=1$ and $n_{LO}=0.65$.}
%\label{fig:V}
        \end{figure*}

The lowest mechanical excitation amplitude with which we could still use the DC rectification method to discern the mechanical resonance peak with this setup is $V_{m}\approx \SI{1.6}{\milli\volt}_{\m{RMS}}$ that corresponds to $\Delta x \approx \SI{3.4e-12}{\meter}$ and $n_{\m{isl}}\approx \SI{0.18}{\elementarycharge}$. The linewidth of the mechanical peak is $\sim 600$ Hz, corresponding to a mechanical Q value of $\sim 15 \times 10^3$. That Q value is modest compared to what is in principle achievable with quartz resonators \cite{goryachev2012}. We expect this is due to the fact that the disk was fully flat in cross section profile, which does not result in confinement of the mechanical vibration in the disk center, and hence there can be pronounced energy leakage through the supports.

\subsection{\label{sec:mix} Homodyne and heterodyne mixing schemes}

Although the DC rectification is a simple and robust method to obtain the signal at large excitation amplitudes, its sensitivity vanishes towards small vibrations. An improved approach is to use the SET as an RF mixer \cite{knobel81} in the setups shown in Figs.~\ref{fig:systems}b,c. Mixer operation entails that a strong local oscillator (LO) at the frequency $\omega_{\m{LO}}$, see  \eref{eq:sinetime}, is applied to the SET gate, while the actuation tone goes to the actuation electrode as above. In homodyne detection, the LO frequency is the same as the mechanical excitation frequency, $\omega_{\m{LO}} = \omega_m$, and the signal appears as a DC offset similar to the  rectification setup. The device has the same bandwidth as the standard SET, but with a tunable measurement center frequency. The DC biasing is done in a  fashion similar to that described in the previous section.  In terms of complexity, the homodyne setup is comparable to the rectification setup. Analogous to \eref{eq:setdcbessel} we obtain the homodyne current
\begin{equation} 
\label{eq:sethomobessel}
\begin{split}
 & I_{\m{DC}}  \simeq  i_0 - \frac{I_0}{2} \cos \LL(\pi n_{g0} \RR) \Big[ J_0(\pi n_{\m{LO}} ) J_0(\pi n_{\m{isl}} ) +
 2 J_{1}(\pi n_{\m{LO}} ) J_{1}(\pi n_{\m{isl}} ) \Big] \,,
\end{split}
\end{equation}
which allows for linear detection at small amplitudes. Here, $J_1$ is the first-order Bessel function of the first kind.  In \fref{fig:9} we see, similar to \fref{fig:6}, the mechanical mode at $f_{0} \approx\SI{8.865}{\mega\hertz}$. In \fref{fig:11} we show the SET current caused by the increase of the mechanical vibration amplitude when the mechanics is actuated on-resonance.
%By comparison of Fig. \ref{fig:8} and \ref{fig:11} it is clear that the homodyne setup has a much steeper dependence of $I_{SET}$ on the resonator displacement. The function resulting from the fit of a linear curve to the data of Fig. \ref{fig:11} has a slope of $\SI{1}{\nano\ampere\per\milli\volt_{RMS}}$ and hence a calibration curve $\frac{\Delta I_{SET}}{\Delta x} = \SI{1.97e15}{\nano\ampere\per\meter}$.
The increased sensitivity of the homodyne detection over the DC method enables the visibility of the mechanical resonance peak at smaller mechanical excitation voltages down to $V_{0}\approx \SI{0.6}{\milli\volt}_{\m{RMS}}$. That corresponds to $\Delta x \approx \SI{1.3e-12}{\meter}$ and $n_{\m{isl}}\approx \SI{0.07}{\elementarycharge}$. The theoretical curve plotted in \fref{fig:11}, calculated from \eref{eq:sethomobessel},  matches well the homodyne  experimental data.

\begin{figure*}
        \begin{subfigure}[IV_het]{0.45\textwidth}
                \includegraphics[width=\linewidth]{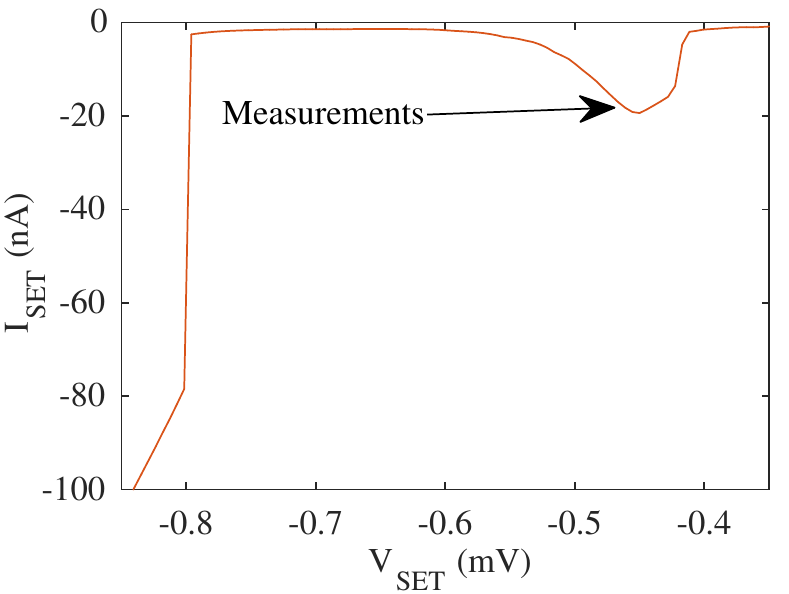}
                \caption{}
                \label{fig:IV_het}
        \end{subfigure}%
        \begin{subfigure}[IV_het_gate]{0.45\textwidth}
                \includegraphics[width=\linewidth]{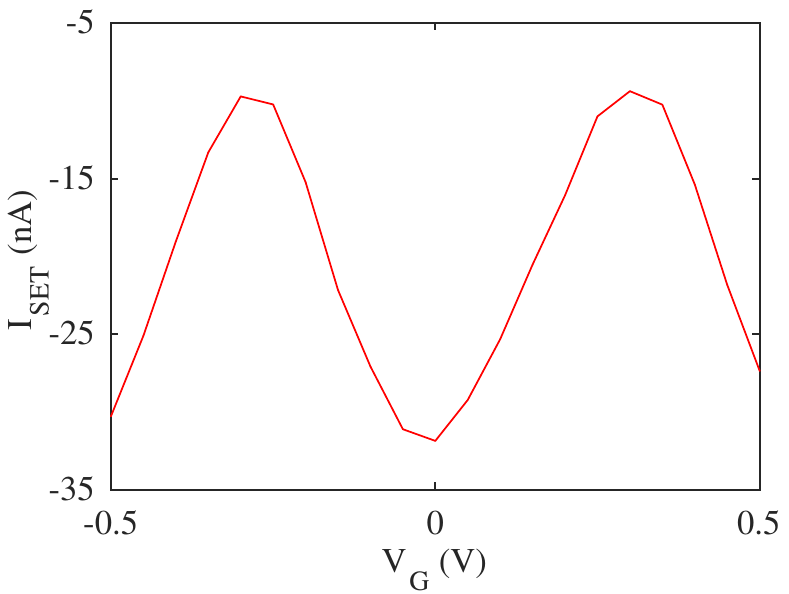}
                \caption{}
                \label{fig:IV_het_gate}
        \end{subfigure}%
        \caption{(\subref{fig:IV_het}) Sample B: DC IV curve for the SET measured with the setup of Fig.~\ref{fig:systems}c when the gate voltage $V_G=0$. (\subref{fig:IV_het_gate}) Dependence of the current through the SET on $V_G$ for the bias values around the JQP peak.}

        \label{fig:IV_het_fig}
        \end{figure*}

Direct conversion to DC, either using the DC method or homodyne detection, encompasses challenges, namely DC offsets from self-mixing of the excitation signal, and in particular, $1/f$ charge noise that is dominant in charge-sensitive devices at low frequencies. In heterodyne detection, the gate electrode is excited by a local oscillator with frequency $\omega_{\m{LO}}$ different from the mechanical excitation frequency $\omega_m$. The response of the SET current is then mixed down to an intermediate frequency (IF) equal to $\omega_{\m{IF}}=\omega_m-\omega_{\m{LO}}$. The relevant current is given as
\begin{equation} 
\label{eq:sethetbessel}
\begin{split}
& I_{\m{IF}} =  I_0 \cos \LL(\pi n_{g0} \RR) J_1 \LL(\pi n_{\m{LO}} \RR) J_1\LL(\pi n_{\m{isl}}\RR)\cos \LL[  (\omega_m-\omega_{\m{LO}}  )t \RR] \,.
\end{split}
\end{equation}
The intermediate frequency needs to be within the bandwidth of the SET setup. Here, we used $\omega_{\m{IF}}/2\pi = 18$ kHz. In the heterodyne setup shown in \fref{fig:systems}c, the SET current is detected by a current pre-amplifier, then acquired by a DAQ board and processed in the frequency domain. Here, we used a different sample B, although the setup is expected to work also with the sample A used for the other measurements discussed.  Figure \ref{fig:IV_het} shows the IV curve of the SET in sample B. We operate at the Josephson Quasiparticle (JQP) peak, originated from the tunnelling of Cooper pairs through one junction followed by two successive quasiparticle tunneling events through the other junction. Instead of the quasiparticle current onset (\fref{fig:2}) used in the measurements discussed above, in this sample we found the JQP peak biasing provides a strong gate modulation as shown in \fref{fig:IV_het_gate}. We relate this to the lower resistance of sample B, providing pronounced Cooper pair tunneling features.

\begin{figure*}
        \begin{subfigure}[6]{0.333\textwidth}
                \includegraphics[width=\linewidth]{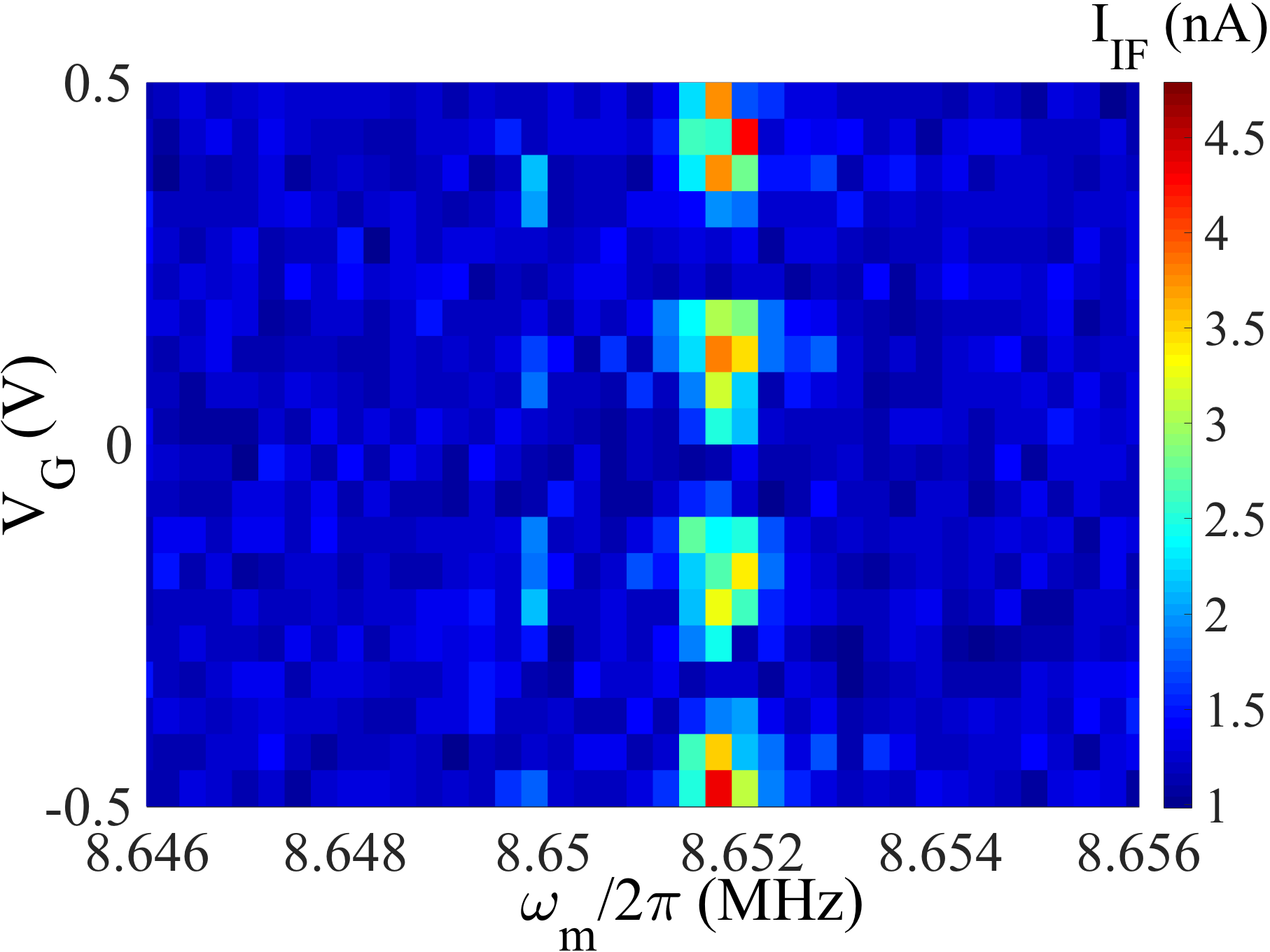}
                \caption{}
                \label{fig:12}
        \end{subfigure}%
        \hfill
        \begin{subfigure}[5]{0.333\textwidth}
                \includegraphics[width=\linewidth]{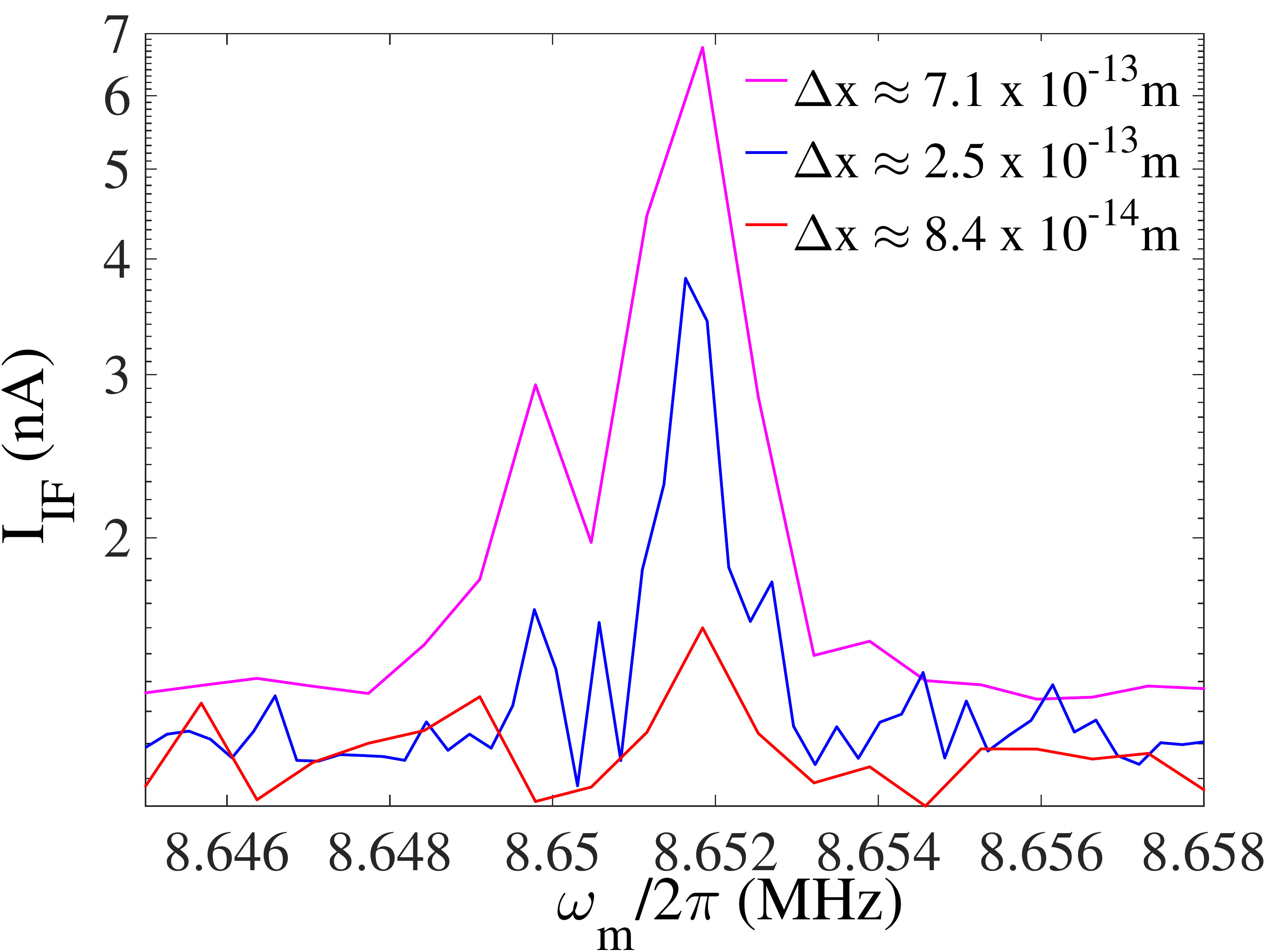}
                \caption{}
                \label{fig:13}
        \end{subfigure}%
        \begin{subfigure}[7]{0.333\textwidth}
                \includegraphics[width=\linewidth]{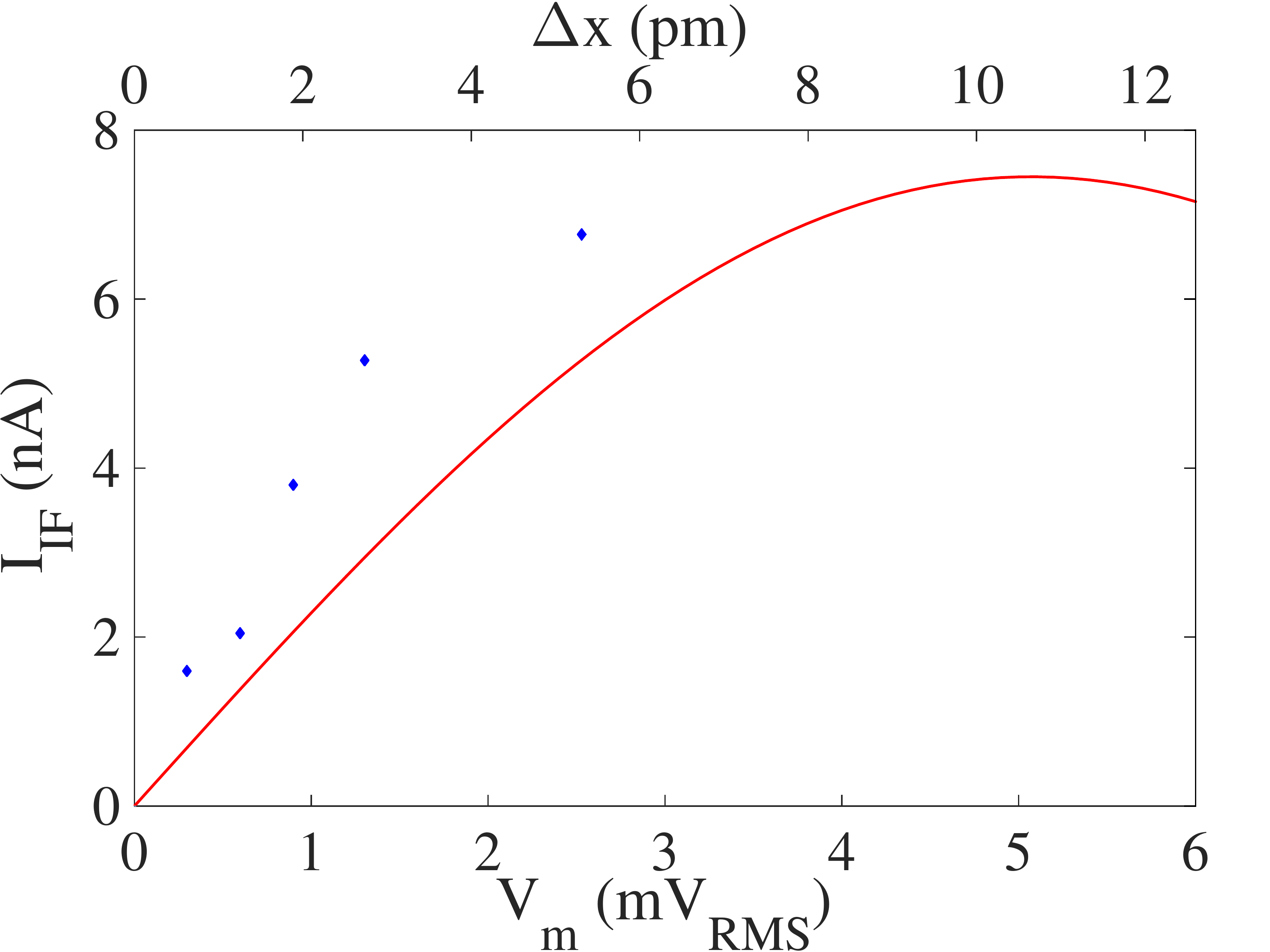}
                \caption{}
                \label{fig:13a}
        \end{subfigure}%
        \caption{Response in  the heterodyne  scheme of  \fref{fig:systems}c (sample B) with the SET bias $V_{\m{SET}} \approx\SI{-0.45}{\milli\volt}$: (\subref{fig:12}) Intermediate-frequency SET current as a function of gate bias and mechanical excitation frequency. The mechanical mode is excited with $V_{m}\approx \SI{0.9}{\milli\volt}_{\m{RMS}}$ and visible at $\approx$ 8.652 MHz; (\subref{fig:13}) Resonance peaks acquired at different mechanical excitation amplitudes; (\subref{fig:13a}) Response when the mechanical resonator is excited on-resonance. The solid line is a theoretical curve based on Eqs.~(\ref{eq:displacement},\ref{eq:charge},\ref{eq:sethetbessel}).}
        \label{fig:V}
        \end{figure*}

Figure \ref{fig:12} displays the detected mechanical signal, with the local oscillator amplitude roughly optimized. %This time, we are detecting another mode in the family of closely spaced peaks at the lowest shear mode. 
Comparing to  \fref{fig:9} it is clear that the heterodyne scheme disposes of the DC response of the SET, hence making the data easier to interpret. Figure \ref{fig:13} shows the mechanical resonance peaks for different mechanical excitation amplitudes. The elimination of low-frequency noise by the heterodyne setup allows to measure the mechanical vibrations down to $\Delta x \approx \SI{6e-13}{\meter}$. This is an improvement  over the homodyne setup, although since the data were obtained from different samples, the origin of the improvement is not necessarily in the mixing scheme used. The dependence of $I_{\m{IF}}$ with the mechanical excitation amplitude is presented in \fref{fig:13a}. The solid line represents a theory curve calculated from \eref{eq:sethetbessel} with $I_0=\SI{22}{\nano\ampere}$ obtained from Fig. \ref{fig:IV_het_gate} and $n_{LO}=0.59$. The data points follow the shape of the theoretical curve, however, display double the amplitude as predicted by independent estimates, a fact we attribute to inaccuracies in the local oscillator amplitude.

%The linear fit to the points of Fig. \ref{fig:13a} yield a slope of $\SI{22.44}{\nano\ampere\per\milli\volt_{RMS}}$ or $\frac{\Delta I_{SET}}{\Delta x} = \SI{4.42e16}{\nano\ampere\per\meter}$.

\subsection{\label{sec:rfset}Radio-frequency SET as a displacement sensor}

In the RF-SET setup \cite{Schoelkopf1238,blencowe2000}, the SET is impedance-matched to 50 $\Omega$ microwave cables via an LC tank resonator circuit. The inductor and capacitor at the input of the SET, see \fref{fig:systems}d, form the tank circuit with resonance frequency $\omega_{LC}=(L C)^{-\frac{1}{2}}$, loaded by the SET.

When the  SET  resistance changes, due to  piezo  vibrations as here, the loading of the LC circuit by the SET changes, hence modulating the reflected power of a monochromatic carrier wave sent down the input line. The carrier appears as a time-dependent voltage across the SET, with the amplitude $\delta V$ \hspace{-2mm}. The SET conductance is 
\begin{equation} 
%\label{eq:sethetbessel}
\begin{split}
& G_{\m{SET}} = \frac{I_{\m{SET}} }{V_{\m{SET}} } = \frac{i_0 - \frac{I_0}{2} \cos \LL(\pi (n_{g0} + n_{\m{isl}} \cos \omega_m t )\RR)}{V_{\m{bias}} + \delta V \cos \omega_{LC} t}  \,.
\end{split}
\end{equation}
%

%
%\begin{equation} 
%\label{eq:sethetbessel}
%\begin{split}
%& G_{\m{SET}} = \frac{I_{\m{SET}} }{V_{\m{SET}} } = \frac{i_0 - \frac{I_0}{2} \cos \LL(\pi (n_{g0} + n_{\m{isl}} \cos \omega t )\RR)}{V_{\m{max}} + \delta V \cos \omega_{LC} t} = \\
%
%& \simeq \LL[ V_{\m{max}} - \delta V \cos \omega_{LC} t\RR] \LL[ i_0 - \frac{I_0}{2} \cos \pi n_{g0} \cos (\pi n_{\m{isl}} \cos \omega t ) + \frac{I_0}{2} \sin \pi n_{g0} \sin ( \pi n_{\m{isl}} \cos \omega t )\RR] = \\
%
%& \Longrightarrow \delta V \frac{I_0}{2}\cos \pi n_{g0}  \cos \omega_{LC} t \cos ( \pi n_{\m{isl}} \cos \omega t ) + \delta V \frac{I_0}{2}\sin \pi n_{g0}  \cos \omega_{LC} t \sin ( \pi n_{\m{isl}} \cos \omega t ) = \\
%
%&= \delta V \frac{I_0}{2}\cos \pi n_{g0}  \cos \omega_{LC} t \LL[ J_0( \pi n_{\m{isl}} ) + 2 \sum_{k=1}^{\infty}
%  (-1)^k J_{2k}( \pi n_{\m{isl}} ) \cos \LL( 2k \omega t \RR)\RR] + \\
%  & +  \delta V \frac{I_0}{2}\sin \pi n_{g0}  \cos \omega_{LC} t \LL[ 2 \sum_{k=0}^{\infty}
%  (-1)^k J_{2k+1}(\pi n_{\m{isl}} ) \cos \LL( (2k + 1) \omega t \RR) \RR]  \\
%\end{split}
%\end{equation}
%
%The first sidebands are
%
%\begin{equation} 
%\label{eq:sethetbessel}
%\begin{split}
%& G_{\m{SET}} =  \delta V \frac{I_0}{2}\sin \pi n_{g0}  \cos \omega_{LC} t \LL[ 2 J_{1}(\pi n_{\m{isl}} ) \cos \LL(  \omega t \RR) \RR]  = \\
%& \delta V \frac{I_0}{2}\sin (\pi n_{g0}  ) J_{1}(\pi n_{\m{isl}} ) \LL[ \cos \LL(  (\omega_{LC} -\omega )t\RR)+\cos \LL(  (\omega_{LC} + \omega )t\RR) \RR]
%\end{split}
%\end{equation}
%
The information is the power of the sidebands that  appear at frequencies offset by the actuation frequency from a carrier applied at $\omega_{LC}$, obtained by detecting the carrier at room temperature. Under $\delta V \ll V$, the first sideband amplitude is
\begin{equation} 
\label{eq:rfsetbessel}
\begin{split}
& I_1 = \delta V \frac{I_0}{2} \LL| \sin (\pi n_{g0}  ) J_{1}(\pi n_{\m{isl}} )  \RR|  \,.
\end{split}
\end{equation}

The transmitted and reflected waves are separated by circulators at base temperature stage of the refrigerator, so the reflected power can be measured by a spectrum analyzer. The bias-T allows for DC biasing the SET, needed to obtain a charge-sensitive response.

\begin{figure*}
     \begin{subfigure}[c]{0.330\textwidth}
     		\includegraphics[width=\linewidth]{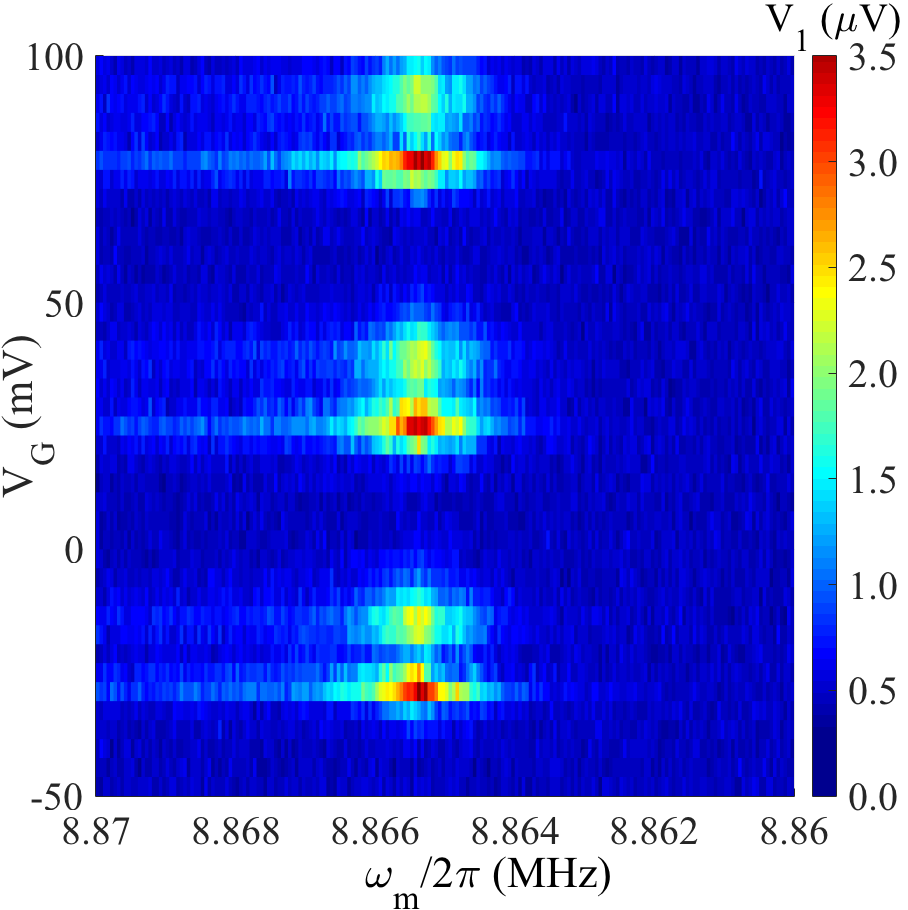}%
                \caption{}
                \label{fig:14a}
     \end{subfigure}%
     \hfill
     \begin{subfigure}[c]{0.33\textwidth}
         \begin{subfigure}{\textwidth}
                \includegraphics[width=\linewidth]{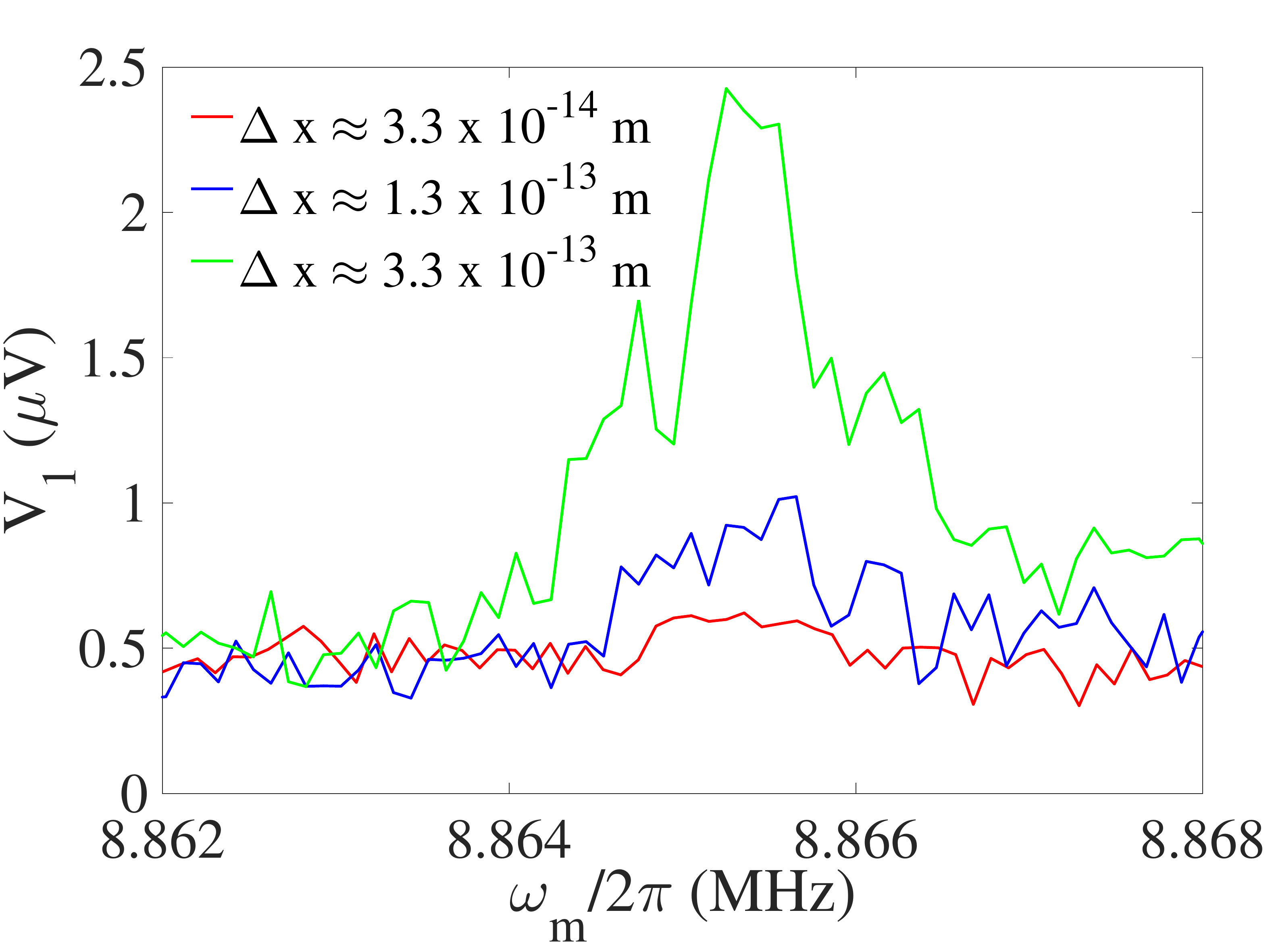}
                \caption{}
                 \label{fig:15}
        \end{subfigure}%
        
        \begin{subfigure}{\textwidth}
                \includegraphics[width=\linewidth]{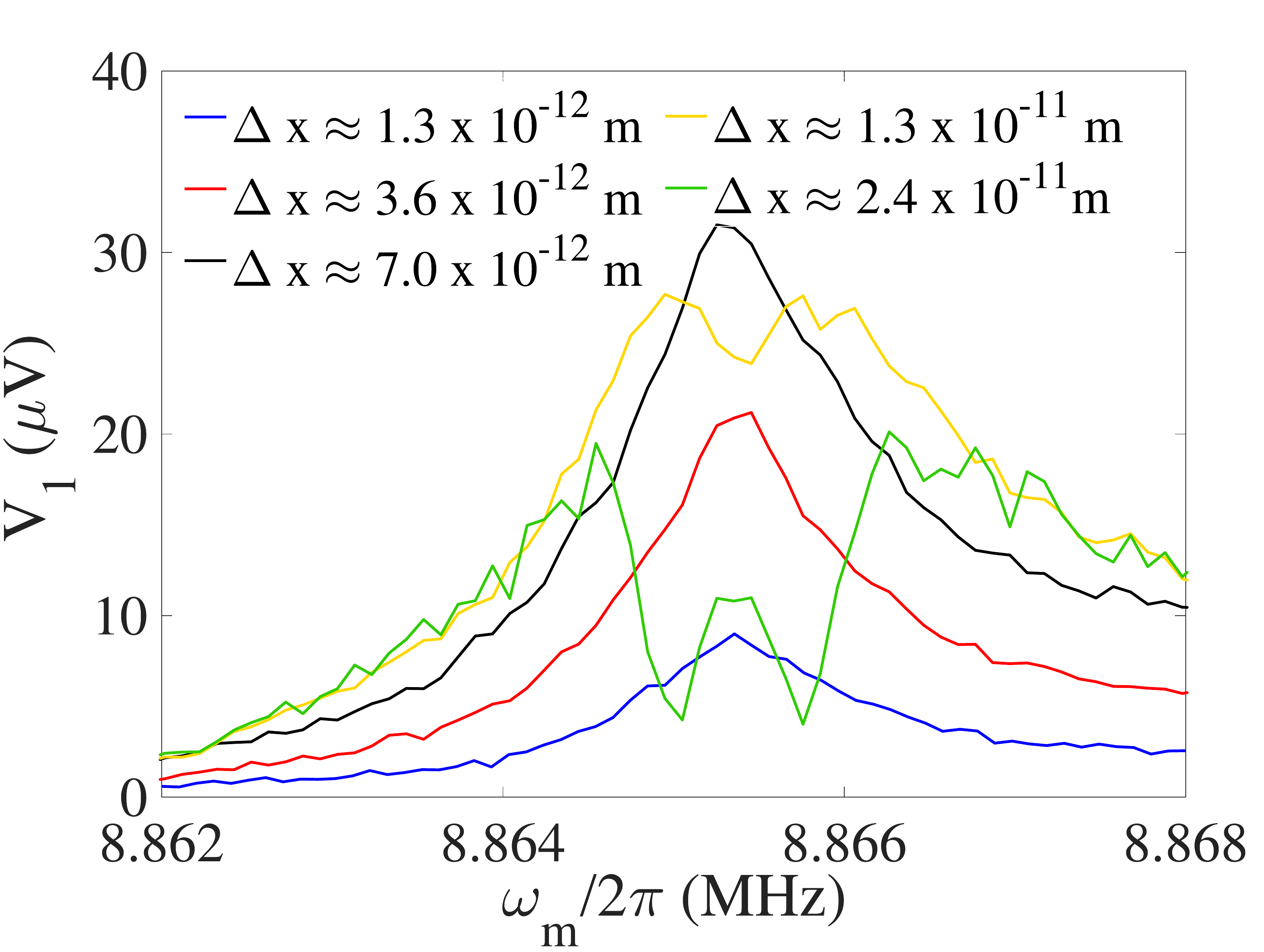}
                \caption{}
                 \label{fig:15a}
        \end{subfigure}%
        \label{fig:15general}
     \end{subfigure}
     \begin{subfigure}[c]{0.3\textwidth}
                   \includegraphics[width=\linewidth]{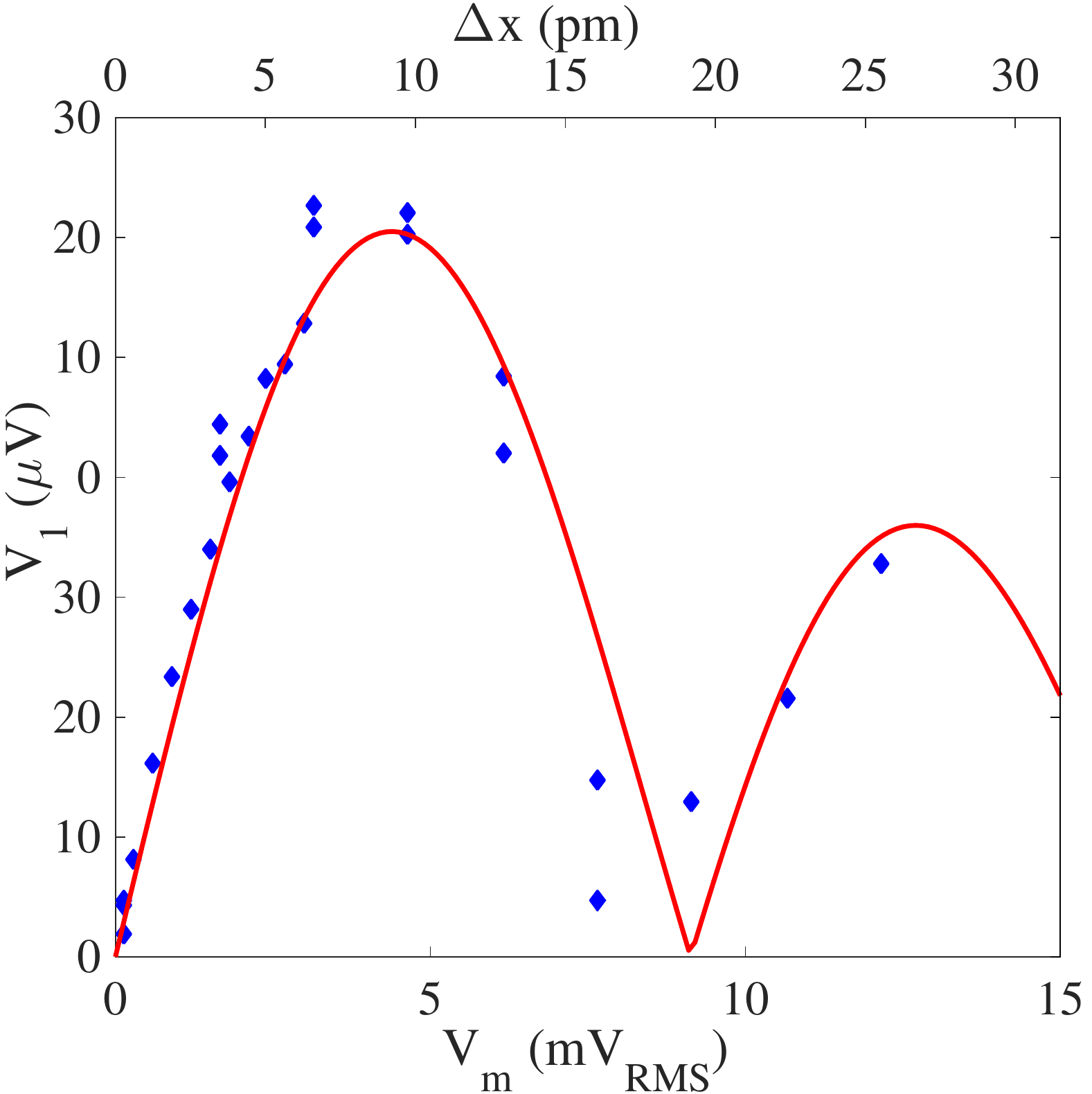}%
                \caption{}
                \label{fig:17}
    \end{subfigure}%
     \caption{Measurement of the quartz disk vibrations using the RF-SET scheme of \fref{fig:systems}d (sample A) biased with $V_{\m{bias}} \approx \SI{-0.18}{\milli\volt}$: (a) Gate bias dependence of the sideband response  whose amplitude is called $V_1$ in this plot; (\subref{fig:15}) Mechanical resonance peaks measured at various low amplitude mechanical excitation voltages, and $V_{G} \approx \SI{40}{\milli\volt}$; (\subref{fig:15a}) As (b), but high amplitudes; (\subref{fig:17}) Response as a function of excitation amplitude with resonant excitation. The solid line is the theory curve from Eqs.~(\ref{eq:displacement},\ref{eq:charge},\ref{eq:rfsetbessel}), with a carrier amplitude of \SI{-86}{\deci\bel m} and an output amplification  of \SI{80}{\deci\bel}.}
                \label{fig:VI}
        \end{figure*}

In the present system the tank circuit capacitance and inductance come from the  bonding wires' and bonding pad's stray inductance and capacitance, respectively. The sample box configuration and bonding wires lengths and placements were simulated with Sonnet software and tweaked to set the tank circuit resonance frequency around \SIrange{4}{5}{\giga\hertz}. As an example,  \fref{fig:14a} shows a RF-SET measurement displaying the mechanical peak. For the following measurements, the gate and SET bias voltages were chosen so that  the maximum SET differential resistance modulation is obtained.

Figs. \ref{fig:15},c present examples of the mechanical resonance curves. Each curve represents the amplitude of the sideband of the carrier. Similar to the other detection methods discussed above, we can enter the nonlinear regime of the SET response, that is, observe the full Bessel dependence in \eref{eq:rfsetbessel}. Splitting of the peaks is due to the fact that  the driven charge $n_{\m{isl}} $ at on-resonance drive is larger than at off-resonant drive. The lowest mechanical excitation amplitude with which we could still discern the mechanical resonance peak with this setup is $V_{m}\approx \SI{15}{\micro \volt}_{\m{RMS}}$, which corresponds to $\Delta x \approx 3 \times 10^{-14}$ m, more than 20 times better than the other techniques.

\section{\label{sec:conc} Conclusions}

We have shown that single electron transistor (SET) is a viable tool to detect minuscule mechanical vibrations in millimetre sized monolithic quartz disk resonators. We compared four detection schemes; DC rectification, homodyne and heterodyne mixing detection, and an approach based on the radio-frequency SET. We found that over the compared schemes, RF-SET is superior in sensitivity over mixing methods,  allowing to measure oscillations down to $\sim 3 \times 10^{-14}$ m within the mechanical linewidth, corresponding to the sensitivity $10^{-15}$ m$/\sqrt{\m{Hz}}$. The zero-point motion of our quartz disk resonators is of the order of $\Delta x_{\m{zp}} \sim \SI{e-19}{\meter}$, hence it is still beyond the capabilities of our SET based detectors. 

The performance can be improved, first of all, by increasing the sensitivity of the rf-SET to the piezo charge. In the current setup, the charge sensitivity is around $\sim 10^{-3}$ $e/\sqrt{\m{Hz}}$ that leaves plenty of room for improvement up to the best demonstrated values $\sim 10^{-6}$ $e/\sqrt{\m{Hz}}$ \cite{Delsing2006}. This entails in particular designing an on-chip tank circuit having a high Q value, and also increasing the charging energy by  fabricating smaller junctions.

Another limiting factor  is that the piezoelectric surface charge  originating from the mechanical strain gets spread across a \SI{6}{\milli\meter} diameter circular area; while our SET sensitive area, the island, is just \SI{3}{\micro\meter} $\times$ \SI{4}{\micro\meter}. This limits the charge coupled to the island to a very small fraction of the total piezoelectric charge generated by a given deformation of the quartz disk. At the moment the amount of charge coupled to the island due to $\Delta x_{\m{zp}}$ is of the order $10^{-8} \, e$.  One could increase the island size in order to boost up the coupled charge. However this would lead to an increase in the island capacitance, decreasing the Coulomb gap and possibly reducing the detector sensitivity. Thus the island size is a trade-off between the SET sensitivity and the charge it couples to. We note, however, that since quartz has a low  dielectric constant $\epsilon_r \sim 4$, the charging energy is currently set by the junctions.

The most important challenge to tackle is to increase the mechanical Q value, allowing to narrow down the detection bandwidth. With high Q values demonstrated by monolithic quartz resonators, combined with the best charge sensitivities mentioned above, and possibly with low-noise Josephson parametric amplifiers to further reduce the detection noise \cite{Siddiqi2015Amp}, one would reach the level of sensitivity needed to observe vibrations at the single-quantum level with the current amount of coupled charge. Plano-convex cross-section profile \cite{onoe2005} of the quartz disks would focus the mode energy in the center, hence mitigating anchor losses that we believe are limiting the losses. 

The work paves the way towards studying massive mechanical resonators near the quantum limit of their motion. Monolithic quartz resonators are suitable for this purpose since they are tangible, sturdy, durable and easily manipulated objects in contrast with other micrometer sized resonators, and  can have high quality factors at frequencies of tens of MHz. They can be easily integrated in other devices without the need of advanced fabrication techniques. The millimetre-size diameter of the disk resonators provides a large sensing area that can be coupled to other macroscopic object to sense mechanical loadings. It can excel at applications that require a simple design that can measure very small strains over a large sensing area.

\begin{acknowledgements}
This work was supported by the Academy of Finland (contract 250280, CoE LTQ, 275245), the European Research Council (615755-CAVITYQPD), the Centre for Quantum Engineering at Aalto University, and by the Finnish Cultural Foundation (Central Fund 00160903). The work benefited from the facilities at the OtaNano - Micronova Nanofabrication Center and at the Low Temperature Laboratory.
\end{acknowledgements}

%\bibliographystyle{spphys} 
%\bibliography{paper_jian}

\begin{thebibliography}{10}
\providecommand{\url}[1]{{#1}}
\providecommand{\urlprefix}{URL }
\expandafter\ifx\csname urlstyle\endcsname\relax
  \providecommand{\doi}[1]{DOI \discretionary{}{}{}#1}\else
  \providecommand{\doi}{DOI \discretionary{}{}{}\begingroup
  \urlstyle{rm}\Url}\fi

\bibitem{slides}
J.A. Sidles, J.L. Garbini, K.J. Bruland, D.~Rugar, O.~Z\"uger, S.~Hoen, C.S.
  Yannoni, Rev. Mod. Phys. \textbf{67}, 249 (1995).
\newblock \urlprefix\url{https://link.aps.org/doi/10.1103/RevModPhys.67.249}

\bibitem{Palomaki710}
T.A. Palomaki, J.D. Teufel, R.W. Simmonds, K.W. Lehnert, Science
  \textbf{342}(6159), 710 (2013).
\newblock \doi{10.1126/science.1244563}.
\newblock \urlprefix\url{http://science.sciencemag.org/content/342/6159/710}

\bibitem{Wollman952}
E.E. Wollman, C.U. Lei, A.J. Weinstein, J.~Suh, A.~Kronwald, F.~Marquardt, A.A.
  Clerk, K.C. Schwab, Science \textbf{349}(6251), 952 (2015).
\newblock \doi{10.1126/science.aac5138}.
\newblock \urlprefix\url{http://science.sciencemag.org/content/349/6251/952}

\bibitem{JuhaSqueeze}
J.M. Pirkkalainen, E.~Damsk\"agg, M.~Brandt, F.~Massel, M.A. Sillanp\"a\"a,
  Phys. Rev. Lett. \textbf{115}, 243601 (2015).
\newblock
  \urlprefix\url{https://link.aps.org/doi/10.1103/PhysRevLett.115.243601}

\bibitem{Lecocq2015}
F.~Lecocq, J.B. Clark, R.W. Simmonds, J.~Aumentado, J.D. Teufel, Phys. Rev. X
  \textbf{5}, 041037 (2015).
\newblock \urlprefix\url{https://link.aps.org/doi/10.1103/PhysRevX.5.041037}

\bibitem{Cleland1996}
A.N. Cleland, M.L. Roukes, Appl. Phys. Lett. \textbf{69}(18), 2653
  (1996).
\newblock \urlprefix\url{http://dx.doi.org/10.1063/1.117548}

\bibitem{Cleland1999256}
A.~Cleland, M.~Roukes, Sensors and Actuators A: Physical \textbf{72}(3), 256
  (1999)

\bibitem{Mohanty2000}
P.~Mohanty, D.A. Harrington, M.L. Roukes, Physica B: Condensed Matter
  \textbf{284-288}, 2143  (2000).
\newblock
  \urlprefix\url{http://www.sciencedirect.com/science/article/pii/S092145269902997X}

\bibitem{Pashkin08}
T.F. Li, Y.A. Pashkin, O.~Astafiev, Y.~Nakamura, J.S. Tsai, H.~Im, Appl. Phys.
  Lett. \textbf{92}, 043112 (2008)

\bibitem{Carr1999}
D.W. Carr, S.~Evoy, L.~Sekaric, H.G. Craighead, J.M. Parpia, Applied Physics
  Letters \textbf{75}(7), 920 (1999).
\newblock \urlprefix\url{http://dx.doi.org/10.1063/1.124554}

\bibitem{Hakseong2017}
H.~Kim, D.H. Shin, K.~McAllister, M.~Seo, S.~Lee, I.S. Kang, B.H. Park, E.E.B.
  Campbell, S.W. Lee, ACS Applied Materials \& Interfaces \textbf{9}(8), 7282
  (2017).
\newblock \urlprefix\url{http://dx.doi.org/10.1021/acsami.6b16278}

\bibitem{vanderZant2009}
G.A. Steele, A.K. H{\"u}ttel, B.~Witkamp, M.~Poot, H.B. Meerwaldt, L.P.
  Kouwenhoven, H.S.J. van~der Zant, Science \textbf{325}(5944), 1103 (2009).
\newblock \urlprefix\url{http://science.sciencemag.org/content/325/5944/1103}

\bibitem{Bachtold2009}
B.~Lassagne, Y.~Tarakanov, J.~Kinaret, D.~Garcia-Sanchez, A.~Bachtold, Science
  \textbf{325}(5944), 1107 (2009).
\newblock \urlprefix\url{http://science.sciencemag.org/content/325/5944/1107}

\bibitem{Cohadon1999}
P.F. Cohadon, A.~Heidmann, M.~Pinard, Phys. Rev. Lett. \textbf{83}, 3174 (1999)

\bibitem{Aspelmeyer2006cool}
S.~Gigan, H.R. B\o"hm, M.~Paternostro, F.~Blaser, G.~Langer, J.B. Hertzberg,
  K.C. Schwab, D.~B\a"uerle, M.~Aspelmeyer, A.~Zeilinger, Nature \textbf{444},
  67 (2006)

\bibitem{Heidmann2006}
O.~Arcizet, P.F. Cohadon, T.~Briant, M.~Pinard, A.~Heidmann, Nature
  \textbf{444}, 71 (2006)

\bibitem{Lehnert2008Nph}
C.A. Regal, J.D. Teufel, K.W. Lehnert, Nature Physics \textbf{4}, 555 (2008)

\bibitem{Flees1997}
D.J. Flees, S.~Han, J.E. Lukens, Phys. Rev. Lett. \textbf{78}, 4817 (1997).
\newblock \urlprefix\url{https://link.aps.org/doi/10.1103/PhysRevLett.78.4817}

\bibitem{Joyez1994}
P.~Joyez, P.~Lafarge, A.~Filipe, D.~Esteve, M.H. Devoret, Phys. Rev. Lett.
  \textbf{72}, 2458 (1994).
\newblock \urlprefix\url{https://link.aps.org/doi/10.1103/PhysRevLett.72.2458}

\bibitem{Knobel2003}
R.G. Knobel, A.N. Cleland, Nature \textbf{424}(6946), 291 (2003).
\newblock \urlprefix\url{http://dx.doi.org/10.1038/nature01773}

\bibitem{LaHaye74}
M.D. LaHaye, O.~Buu, B.~Camarota, K.C. Schwab, Science \textbf{304}(5667), 74
  (2004).
\newblock \doi{10.1126/science.1094419}.
\newblock \urlprefix\url{http://science.sciencemag.org/content/304/5667/74}

\bibitem{Nakamura2010nems}
Y.A. Pashkin, T.F. Li, J.P. Pekola, O.~Astafiev, D.A. Knyazev, F.~Hoehne,
  H.~Im, Y.~Nakamura, J.S. Tsai, Appl. Phys. Lett. \textbf{96}(26),
  263513 (2010).
\newblock \urlprefix\url{http://dx.doi.org/10.1063/1.3455880}

\bibitem{LaHaye2009}
M.D. LaHaye, J.~Suh, P.M. Echternach, K.C. Schwab, M.L. Roukes, Nature
  \textbf{459}, 960 (2009)

\bibitem{ClelandMartinis}
A.D. O'Connell, M.~Hofheinz, M.~Ansmann, R.C. Bialczak, M.~Lenander, E.~Lucero,
  M.~Neeley, D.~Sank, H.~Wang, M.~Weides, J.~Wenner, J.M. Martinis, A.N.
  Cleland, Nature \textbf{464}, 697 (2010)

\bibitem{transmonnems}
J.M. Pirkkalainen, S.U. Cho, J.~Li, G.S. Paraoanu, P.J. Hakonen, M.A.
  Sillanp{\"a}{\"a}, Nature \textbf{494}, 211 (2013)

\bibitem{LSETNEMSexp}
J.M. Pirkkalainen, S.U. Cho, F.~Massel, J.~Tuorila, T.T. Heikkila, P.J.
  Hakonen, M.A. Sillanp\"a\"a, Nature Communications \textbf{6}, 6981 (2015)

\bibitem{Delsing2014}
M.V. Gustafsson, T.~Aref, A.F. Kockum, M.K. Ekstr\"om, G.~Johansson,
  P.~Delsing, Science \textbf{346}, 207 (2014)

\bibitem{LaHaye2016}
F.~Rouxinol, Y.~Hao, F.~Brito, A.O. Caldeira, E.K. Irish, M.D. LaHaye,
  Nanotechnology \textbf{27}(36), 364003 (2016).
\newblock \urlprefix\url{http://stacks.iop.org/0957-4484/27/i=36/a=364003}

\bibitem{Santos2017}
J.T. Santos, J.~Li, J.~Ilves, C.F. Ockeloen-Korppi, M.~Sillanp\"a\"a, New
  Journal of Physics \textbf{19}(10), 103014 (2017).
\newblock \urlprefix\url{http://stacks.iop.org/1367-2630/19/i=10/a=103014}

\bibitem{woolley2016}
M.~Woolley, M.~Emzir, G.~Milburn, M.~Jerger, M.~Goryachev, M.~Tobar,
  A.~Fedorov, Phys.~Rev.~B \textbf{93}, 224518 (2016)

\bibitem{Knobel2002}
R.~Knobel, A.N. Cleland, Appl. Phys. Lett. \textbf{81}(12), 2258 (2002).
\newblock \urlprefix\url{http://dx.doi.org/10.1063/1.1507616}

\bibitem{Sidles91}
J.A. Sidles, Appl. Phys. Lett. \textbf{58}(24), 2854 (1991).
\newblock \urlprefix\url{https://doi.org/10.1063/1.104757}

\bibitem{Pettersson96}
J.~Pettersson, P.~Wahlgren, P.~Delsing, D.B. Haviland, T.~Claeson, N.~Rorsman,
  H.~Zirath, Phys. Rev. B \textbf{53}, R13272 (1996).
\newblock \urlprefix\url{https://link.aps.org/doi/10.1103/PhysRevB.53.R13272}

\bibitem{Visscher96}
E.H. Visscher, J.~Lindeman, S.M. Verbrugh, P.~Hadley, J.E. Mooij, W.~van~der
  Vleuten, Appl. Phys. Lett. \textbf{68}(14), 2014 (1996).
\newblock \urlprefix\url{http://dx.doi.org/10.1063/1.115622}

\bibitem{knobel81}
R.~Knobel, C.S. Yung, A.N. Cleland, Appl. Phys. Lett. \textbf{81}(3), 532
  (2002).
\newblock \urlprefix\url{http://dx.doi.org/10.1063/1.1493221}

\bibitem{Abuelmaatti2009}
M.T. Abuelma'atti, Analog Integrated Circuits and Signal Processing
  \textbf{61}(3), 223 (2009).
\newblock \urlprefix\url{https://doi.org/10.1007/s10470-009-9304-z}

\bibitem{Schoelkopf1238}
R.J. Schoelkopf, P.~Wahlgren, A.A. Kozhevnikov, P.~Delsing, D.E. Prober,
  Science \textbf{280}(5367), 1238 (1998).
\newblock \urlprefix\url{http://science.sciencemag.org/content/280/5367/1238}

\bibitem{Devoret2000}
M.H. Devoret, R.J. Schoelkopf, Nature \textbf{406}(6799), 1039 (2000).
\newblock \urlprefix\url{http://dx.doi.org/10.1038/35023253}

\bibitem{goryachev2012}
M.~Goryachev, D.L. Creedon, E.N. Ivanov, S.~Galliou, R.~Bourquin, M.E. Tobar,
  Appl.~Phys.~Lett. \textbf{100}(24), 243504 (2012)

\bibitem{blencowe2000}
M.~Blencowe, M.~Wybourne, Appl. Phys. Lett. \textbf{77}, 3845 (2000)

\bibitem{Delsing2006}
H.~Brenning, S.~Kafanov, T.~Duty, S.~Kubatkin, P.~Delsing, J. Appl.
  Phys. \textbf{100}(11), 114321 (2006).
\newblock \urlprefix\url{http://aip.scitation.org/doi/abs/10.1063/1.2388134}

\bibitem{Siddiqi2015Amp}
C.~Macklin, K.~OÕBrien, D.~Hover, M.E. Schwartz, V.~Bolkhovsky, X.~Zhang, W.D.
  Oliver, I.~Siddiqi, Science \textbf{350}, 307 (2015)

\bibitem{onoe2005}
M.~Onoe, in \emph{Proceedings of the 2005 IEEE International Frequency Control
  Symposium and Exposition, 2005.} (IEEE, 2005), pp. 433--441

\end{thebibliography}

\end{document}